\documentclass[iop]{emulateapj}

\usepackage{graphicx}
\def\dlambda{$\lambda\lambda$}
\def\kms{km~s$^{-1}$} 

\slugcomment{Draft version \today}

\shorttitle{Multi-wavelength Observations of SN 2011ei}
\shortauthors{Milisavljevic et al.}

\newenvironment{my_enumerate}{
\begin{enumerate}
        \setlength{\leftmargin}{0pt}
        \setlength{\itemsep}{1pt}
        \setlength{\parskip}{2pt}
	\setlength{\parsep}{1pt}}{\end{enumerate}
}

\begin{document}

\def\cfa{1}
\def\uab{2}
\def\nrao{3}
\def\dartmouth{4}
\def\lco{5}
\def\saao{6}
\def\salt{7}
\def\uw{8}
\def\ugoe{9}
\def\york{10}
\def\hart{11}
\def\su{12}
\def\mp{13}
\def\inafpad{14}
\def\kavli{15}
\def\inaf{16}
\def\scuola{17}
\def\bonn{18}
\def\tata{19}
\def\cu{20}
\def\nyu{21}
\def\uchile{22}
\def\inafpad{23}
\def\unc{24}

\title{Multi-wavelength Observations of Supernova 2011\lowercase{ei}:
  Time-Dependent Classification of Type II\lowercase{b} and
  I\lowercase{b} Supernovae and Implications for their Progenitors}

\author{Dan~Milisavljevic\altaffilmark{\cfa},
        Raffaella~Margutti\altaffilmark{\cfa},
        Alicia~M.~Soderberg\altaffilmark{\cfa},                
        Giuliano~Pignata\altaffilmark{\uab}, 	 
        Laura~Chomiuk\altaffilmark{\nrao,\cfa},   	 
        Robert~A.~Fesen\altaffilmark{\dartmouth},
        Filomena~Bufano\altaffilmark{\uab},
        Nathan~E.~Sanders\altaffilmark{\cfa},         
        Jerod~T.~Parrent\altaffilmark{\dartmouth,\lco},
        Stuart~Parker,
        Paolo~Mazzali\altaffilmark{\mp,\inafpad,\kavli},
        Elena~Pian\altaffilmark{\inaf,\scuola,\kavli}
        Timothy~Pickering\altaffilmark{\saao,\salt},
        David~A.~H.~Buckley\altaffilmark{\saao,\salt},
        Steven~M.~Crawford\altaffilmark{\saao,\salt},
        Amanda~A.~S.~Gulbis\altaffilmark{\saao,\salt},
        Christian~Hettlage\altaffilmark{\saao,\salt},
        Eric~Hooper\altaffilmark{\uw},
        Kenneth~H.~Nordsieck\altaffilmark{\uw},
        Darragh~O'Donoghue\altaffilmark{\salt},
        Tim-Oliver~Husser\altaffilmark{\ugoe},
        Stephen~Potter\altaffilmark{\saao}, 
        Alexei~Kniazev\altaffilmark{\saao,\salt},
        Paul~Kotze\altaffilmark{\saao,\salt},
        Encarni~Romero-Colmenero\altaffilmark{\saao,\salt},
        Petri~Vaisanen\altaffilmark{\saao,\salt},
        Marsha~Wolf\altaffilmark{\uw},
        Michael~F.~Bietenholz\altaffilmark{\york,\hart}
        Norbert~Bartel\altaffilmark{\york},
        Claes~Fransson\altaffilmark{\su},
        Emma~S.~Walker\altaffilmark{\scuola},
        Andreas~Brunthaler\altaffilmark{\bonn,\nrao},
        Sayan~Chakraborti\altaffilmark{\tata},
        Emily~M.~Levesque\altaffilmark{\cu},
        Andrew~MacFadyen\altaffilmark{\nyu},
        Colin~Drescher,
        Greg~Bock,
        Peter~Marples,
        Joseph~P.~Anderson\altaffilmark{\uchile},
        Stefano~Benetti\altaffilmark{\inafpad},
        Daniel~Reichart\altaffilmark{\unc}, and
        Kevin~Ivarsen\altaffilmark{\unc}
}
\altaffiltext{\cfa}{Harvard-Smithsonian Center for Astrophysics, 60 Garden Street,
                 Cambridge, MA, 02138. Electronic address: dmilisav@cfa.harvard.edu} 
\altaffiltext{\uab}{Departamento de Ciencias Fisicas, Universidad Andres Bello, Avda. Republica 252, Santiago, Chile}
\altaffiltext{\nrao}{National Radio Astronomy Observatory, P.O. Box O, Socorro, NM 87801, USA}
\altaffiltext{\dartmouth}{Department of Physics \& Astronomy, Dartmouth
                 College, 6127 Wilder Lab, Hanover, NH, 03755}
\altaffiltext{\lco}{Las Cumbres Observatory Global Telescope Network, Goleta, CA, US}
\altaffiltext{\saao}{South African Astronomical Observatory, PO Box 9,
Observatory 7935, Cape Town, South Africa}
\altaffiltext{\salt}{Southern African Large Telescope, PO Box 9,
Observatory 7935, Cape Town, South Africa}
\altaffiltext{\uw}{Department of Astronomy, University of Wisconsin, Madison, WI
                  53706, USA}
\altaffiltext{\ugoe}{Institut f\"ur Astrophysik,
        Georg-August-Universit\"at, Friedrich-Hund-Platz 1, 37077 G\"ottingen, Germany}
\altaffiltext{\york}{Department of Physics and Astronomy, York University,
  Toronto, Ontario M3J 1P3, Canada}
\altaffiltext{\hart}{Hartebeesthoek Radio Observatory, P.O. Box 443, Krugersdorp, 1740, South Africa}
\altaffiltext{\su}{Department of Astronomy, The Oskar Klein Centre,
  Stockholm University, 106 91 Stockholm, Sweden}
\altaffiltext{\mp}{Max-Planck-Institut f\"ur Astrophysik,
  Karl-Schwarzschild-Strasse 1, 85748 Garching, Germany}
\altaffiltext{\kavli}{Kavli Institute for Theoretical Physics, Kohn
        Hall, University of California at Santa Barbara, Santa
        Barbara, CA 93106-4030}
\altaffiltext{\inaf}{INAF, Trieste Astronomical Observatory, Via G.B. Tiepolo 11, I-34143 Trieste, Italy}
\altaffiltext{\scuola}{Scuola Normale Superiore, Piazza dei Cavalieri 7, 56126 Pisa, Italy;
}
\altaffiltext{\bonn}{Max-Planck-Institut f\"ur Radioastronomie, Auf dem H\"ugel 69, 53121 Bonn, Germany}
\altaffiltext{\tata}{Department of Astronomy and Astrophysics, Tata Institute of Fundamental Research, 1 Homi Bhabha Road, Mumbai 400 005, India}
\altaffiltext{\cu}{CASA, Department of Astrophysical and Planetary
        Sciences, University of Colorado, 389-UCB, Boulder, CO 80309, USA} 
\altaffiltext{\nyu}{Department of Physics, New York University, 4
  Washington Place, New York, NY 10003}
\altaffiltext{\uchile}{Departamento de Astronomia, Universidad de Chile, Casilla 36-D, Santiago, Chile}
\altaffiltext{\inafpad}{INAF - Osservatorio Astronomico di Padova, Vicolo dell'Osservatorio 5, I-35122, Padova, Italy}
\altaffiltext{\unc}{Department of Physics and Astronomy, University of North Carolina at Chapel Hill, Campus Box 3255, Chapel Hill, NC 27599-3255, USA}

\begin{abstract}

  We present X-ray, UV/optical, and radio observations of the
  stripped-envelope, core-collapse supernova (SN) 2011ei, one of the
  least luminous SNe IIb or Ib observed to date. Our observations
  begin with a discovery within $\sim 1$ day of explosion and span
  several months afterward. Early optical spectra exhibit broad, Type
  II-like hydrogen Balmer profiles that subside rapidly and are
  replaced by Type Ib-like He-rich features on the timescale of one
  week. High-cadence monitoring of this transition suggests that
  absorption attributable to a high velocity ($\ga 12,000$ \kms)
  H-rich shell is not rare in Type Ib events. Radio observations imply
  a shock velocity of $v \approx 0.13\,c$ and a progenitor star
  mass-loss rate of $\dot{M}\approx 1.4 \times 10^{-5}\;M_{\odot}\;\rm
  yr^{-1}$ (assuming wind velocity $v_w=10^3~\rm km~s^{-1}$). This is
  consistent with independent constraints from deep X-ray observations
  with \emph{Swift}-XRT and \emph{Chandra}.  Overall, the
  multi-wavelength properties of SN\,2011ei are consistent with the
  explosion of a lower-mass ($3-4\;M_{\odot}$), compact ($R_* \lesssim
  1 \times 10^{11}$~cm), He core star. The star retained a thin
  hydrogen envelope at the time of explosion, and was embedded in an
  inhomogeneous circumstellar wind suggestive of modest episodic
  mass-loss.  We conclude that SN\,2011ei's rapid spectral
  metamorphosis is indicative of time-dependent classifications that
  bias estimates of explosion rates for Type IIb and Ib objects, and
  that important information about a progenitor star's evolutionary
  state and mass-loss immediately prior to SN explosion can be
  inferred from timely multi-wavelength observations.

\end{abstract}



\section{Introduction}
\label{sec:Intro}

\setcounter{footnote}{0}

Core-collapse supernovae (SNe) are traditionally defined by their
spectroscopic properties in the optical (see \citealt{Filippenko97}
for review).  SNe Type II show hydrogen lines at all epochs. SNe Type
Ib do not show hydrogen lines, but do show conspicuous He features,
and SNe Type Ic show neither H nor He. These classifications are
thought to reflect compositional differences in the envelopes of the
massive progenitor stars immediately prior to explosion. The envelopes
of SNe Ib are H-deficient while those of SNe Ic are both H- and
He-deficient.

However, exceptions and transitional subtypes exist that blur these
formal classifications. Observations of SN 1987K \citep{Filippenko88}
showed a gradual transition in its optical spectra from Type II-like
lines at photospheric epochs ($\la 2$ months post-outburst) to Type
Ib-like features at nebular epochs ($\ga 5$ months). This remarkable
spectroscopic evolution demonstrated the existence of an intermediary
between the H-rich Type II and H-deficient Type Ib and Ic classes
(hereafter Type Ibc). SN 1993J \citep{Filippenko93} was a well-studied
additional example of this breaching of SN classifications and has
come to be thought of as a prototype of SNe Type IIb
\citep{Woosley94}. Like the SNe Ibc objects, SNe IIb are believed to
be massive stars but only partially stripped of their outer H-rich
envelopes.

\begin{deluxetable*}{lcccccccc}[htp!]
\footnotesize
\centering
\tablecaption{Combined \emph{Swift}-UVOT and PROMPT photometry of SN\,2011ei}
\tablecolumns{9}
\tablewidth{0pt}
\tablehead{\colhead{JD}                          &
           \multicolumn{4}{c}{\emph{Swift}-UVOT} &
	   \multicolumn{4}{c}{PROMPT}            \\
           \colhead{$-2400000$} &
           \colhead{$uvw1$}     &
           \colhead{$u$}        &
           \colhead{$b$}        &
           \colhead{$v$}        &
           \colhead{$B$}        &
           \colhead{$V$}        &
           \colhead{$R$}        &
           \colhead{$I$}}
\startdata
  55769.73  & \nodata      & \nodata      & \nodata      & \nodata      & 
              19.264 0.058 & 18.786 0.104 & \nodata      & \nodata      \\
  55770.65  & \nodata      & \nodata      & \nodata      & \nodata      &
              \nodata      & 18.445 0.065 & 18.087 0.024 & 18.127 0.049 \\
  55776.82  & \nodata      & 16.67 0.06   & 17.63 0.07   & 17.58 0.11   &
              \nodata      & \nodata      & \nodata      & \nodata      \\ 
  55776.84  & 17.95 0.07   & \nodata      & \nodata      & \nodata      &
              \nodata      & \nodata      & \nodata      & \nodata      \\
  55777.71  & 18.31 0.12   & 16.75 0.09   & 17.86 0.11   & \nodata      &
              \nodata      & \nodata      & \nodata      & \nodata      \\ 
  55778.52  & 17.90 0.13   & \nodata      & \nodata      & 17.53 0.21   &
              \nodata      & \nodata      & \nodata      & \nodata      \\ 
  55779.25  & \nodata      & 16.67 0.07   & 17.44 0.08   & \nodata      &
              \nodata      & \nodata      & \nodata      & \nodata      \\
  55779.32  & 17.82 0.09   & \nodata      & \nodata      & 17.41 0.15   &
              \nodata      & \nodata      & \nodata      & \nodata      \\ 
  55780.64  & 17.91 0.10   & 16.61 0.08   & 17.38 0.08   & 17.31 0.14   &
              \nodata      & \nodata      & \nodata      & \nodata      \\ 
  55782.61  & 17.90 0.10   & \nodata      & \nodata      & 17.10 0.12   &
              \nodata      & \nodata      & \nodata      & \nodata      \\ 
  55783.14  & \nodata      & 16.70 0.07   & 17.38 0.08   & \nodata      &
              \nodata      & \nodata      & \nodata      & \nodata      \\
  55783.75  & 18.04 0.16   & \nodata      & \nodata      & \nodata      &
              \nodata      & \nodata      & \nodata      & \nodata      \\
  55783.77  & \nodata      & \nodata      & \nodata      & \nodata      &
              17.381 0.020 & 17.031 0.013 & 16.866 0.011 & 16.729 0.019\\  
  55784.68  & \nodata      & \nodata      & \nodata      & \nodata      &
              17.362 0.017 & 17.020 0.017 & 16.846 0.012 & 16.695 0.021\\
  55785.58  & \nodata      & \nodata      & \nodata      & \nodata      &
              17.422 0.019 & 17.006 0.025 & 16.813 0.022 & \nodata     \\
  55788.13  & 18.31 0.11   & 17.27 0.10   & 17.48 0.09   & 16.99 0.11   &
              \nodata      & \nodata      & \nodata      & \nodata      \\ 
  55788.70  & \nodata      & \nodata      & \nodata      & \nodata      &
              17.636 0.018 & 17.006 0.019 & 16.796 0.013 & 16.575 0.021 \\
  55788.96  & 18.40 0.12   & 17.44 0.11   & 17.48 0.09   & \nodata      &
              \nodata      & \nodata      & \nodata      & \nodata      \\
  55789.30  & \nodata      & \nodata      & \nodata      & 17.07 0.10
  &           \nodata      & \nodata      & \nodata      & \nodata    \\
  55789.66  & 18.56 0.13   & 17.55 0.12   & 17.49 0.09   & \nodata      &
              \nodata      & \nodata      & \nodata      & \nodata      \\ 
  55791.03  & 18.49 0.13   & 17.64 0.13   & 17.61 0.09   & \nodata      &
              \nodata      & \nodata      & \nodata      & \nodata      \\ 
  55791.44  & \nodata      & \nodata      & \nodata      & 17.17 0.11  &
              \nodata      & \nodata      & \nodata      & \nodata     \\
  55792.04  & 18.43 0.12   & \nodata      & \nodata      & \nodata      &
              \nodata      & \nodata      & \nodata      & \nodata      \\
  55792.23  & \nodata      & 18.12 0.15   & 17.84 0.10   & \nodata      &
              \nodata      & \nodata      & \nodata      & \nodata      \\ 
  55792.63  & 18.74 0.23   & \nodata      & \nodata      & \nodata      &
              \nodata      & \nodata      & \nodata      & \nodata      \\
  55794.01  & 18.61 0.18   & \nodata      & \nodata      & \nodata      &
              \nodata      & \nodata      & \nodata      & \nodata      \\
  55794.48  & \nodata      & 18.70 0.23   & 18.15 0.12   & \nodata      &
              \nodata      & \nodata      & \nodata      & \nodata      \\ 
  55794.59  & \nodata      & \nodata      & \nodata      & \nodata      &
              18.345 0.031 & 17.354 0.019 & 16.950 0.019 & 16.626 0.026 \\ 
  55795.18  & 18.93 0.16   & \nodata      & \nodata      & \nodata      &
              \nodata      & \nodata      & \nodata      & \nodata      \\
  55795.73  & \nodata      & \nodata      & \nodata      & \nodata      &
              18.508 0.038 & 17.497 0.027 & 16.999 0.018 & 16.674 0.022 \\
  55796.13  & \nodata      & 19.12 0.29   & 18.41 0.15   & \nodata      &
              \nodata      & \nodata      & \nodata      & \nodata      \\ 
  55796.59  & 18.94 0.12   & \nodata      & \nodata      & \nodata      &
              \nodata      & \nodata      & \nodata      & \nodata      \\
  55798.09  & \nodata      & \nodata      & 18.61 0.16   & \nodata      &
              \nodata      & \nodata      & \nodata      & \nodata      \\ 
  55798.72  & \nodata      & \nodata      & \nodata      & \nodata      &
              18.915 0.054 & 17.721 0.022 & 17.085 0.015 & 16.765 0.019 \\
  55800.13  & 18.95 0.12   & 19.59 0.43   & \nodata      & \nodata      &
              \nodata      & \nodata      & \nodata      & \nodata      \\
  55802.11  & \nodata      & \nodata      & 18.92 0.19   & \nodata      &
              \nodata      & \nodata      & \nodata      & \nodata      \\ 
  55802.74  & \nodata      & \nodata      & \nodata      & \nodata      &
              19.131 0.242 & 17.980 0.034 & 17.210 0.017 & 16.846 0.018 \\
  55804.04  & 19.04 0.12   & 20.05 0.63   & \nodata      & \nodata      &
              \nodata      & \nodata      & \nodata      & \nodata      \\
  55805.77  & \nodata      & \nodata      & \nodata      & \nodata      &
              \nodata      & 18.185 0.023 & 17.297 0.017 & 16.973 0.017 \\
  55805.98  & \nodata      & \nodata      & 19.27 0.25   & \nodata      &
              \nodata      & \nodata      & \nodata      & \nodata      \\ 
  55808.15  & \nodata      & 19.98 0.67   & \nodata      & \nodata      &
              \nodata      & \nodata      & \nodata      & \nodata      \\
  55819.61  & \nodata      & \nodata      & \nodata      & \nodata      &
              \nodata      & 18.337 0.055 & 17.552 0.020 & 17.167 0.024 
\enddata

\tablecomments{Uncertainties are adjacent to measurements and are at
  the 68\% confidence level.}

\label{tab:phot}
\end{deluxetable*}

\begin{deluxetable*}{lcccccccc}[htp!]
\footnotesize
\centering
\tablecaption{PROMPT photometry of the local sequence stars in the field of NGC 6925}
\tablecolumns{7}
\tablewidth{0pt}
\tablehead{\colhead{No.}          &
           \colhead{RA(J2000.0)}  &
           \colhead{DEC(J2000.0)} &
           \colhead{$B$}          &
           \colhead{$V$}          &
           \colhead{$R$}          &
           \colhead{$I$}}
\startdata
 1 & 20:34:31.425 &  -31:55:37.13 &    15.350     0.023  &   14.658     0.033  &   14.268     0.036   &    13.898     0.081 \\
 2 & 20:34:28.104 &  -31:55:54.40 &    17.904     0.031  &   17.183     0.027  &   16.758     0.025   &    16.380     0.025 \\
 3 & 20:34:26.771 &  -31:57:32.56 &    15.680     0.015  &   14.989     0.027  &   14.574     0.037   &    14.220     0.048 \\
 4 & 20:34:37.298 &  -31:59:10.83 &    16.843     0.012  &   16.105     0.034  &   15.666     0.044   &    15.235     0.048 \\
 5 & 20:34:35.922 &  -31:59:48.73 &    17.746     0.023  &   17.335     0.031  &   17.027     0.045   &    16.724     0.046 \\
 6 & 20:34:37.883 &  -32:00:03.00 &    17.721     0.028  &   17.053     0.031  &   16.660     0.017   &    16.283     0.029 \\
 7 & 20:34:35.853 &  -32:00:08.29 &    17.139     0.057  &   16.464     0.039  &   16.041     0.028   &    15.682     0.062 \\
 8 & 20:34:35.118 &  -32:00:55.90 &    17.349     0.029  &   16.411     0.024  &   15.858     0.048   &    15.404     0.064 \\
 9 & 20:34:32.157 &  -32:01:06.60 &    17.927     0.011  &   17.396     0.035  &   17.065     0.033   &    16.720     0.046 \\
10 & 20:34:32.166 &  -32:01:22.16 &    17.658     0.062  &   17.061     0.052  &   16.690     0.040   &    16.344     0.046 \\
11 & 20:34:32.923 &  -32:01:43.05 &    18.444     0.020  &   17.061     0.018  &   16.129     0.046   &    15.359     0.078 \\
12 & 20:34:33.090 &  -32:02:17.52 &    17.200     0.054  &   16.363     0.034  &   15.877     0.045   &    15.429     0.063 \\
13 & 20:34:31.201 &  -32:02:05.58 &    18.535     0.158  &   17.278     0.037  &   16.383     0.062   &    15.722     0.063 \\
14 & 20:34:31.086 &  -32:03:03.46 &    18.073     0.053  &   17.483     0.039  &   17.105     0.054   &    16.738     0.024 \\
15 & 20:34:25.617 &  -32:02:06.45 &    17.324     0.052  &   16.661     0.033  &   16.249     0.049   &    15.902     0.072 \\
16 & 20:34:26.544 &  -32:00:29.45 &    17.576     0.026  &   16.749     0.036  &   16.264     0.046   &    15.835     0.073 \\
17 & 20:34:24.061 &  -32:00:48.81 &    16.666     0.045  &   15.924     0.044  &   15.477     0.046   &    15.094     0.064 \\
18 & 20:34:23.184 &  -32:01:34.76 &    18.284     0.047  &   17.222     0.044  &   16.590     0.033   &    16.110     0.072 \\
19 & 20:34:19.306 &  -32:02:19.59 &    15.962     0.029  &   15.272     0.038  &   14.852     0.050   &    14.492     0.070 \\
20 & 20:34:18.786 &  -32:01:26.81 &    18.255     0.012  &   17.336     0.064  &   16.765     0.069   &    16.343     0.033 \\
21 & 20:34:14.755 &  -32:03:33.72 &    18.784     1.180  &   17.378     0.044  &   16.967     0.047   &    16.621     0.085 \\
22 & 20:34:12.242 &  -32:03:10.87 &    16.758     0.043  &   15.789     0.024  &   15.227     0.051   &    14.758     0.075 \\
23 & 20:34:08.874 &  -32:02:50.79 &    15.787     0.016  &   14.658     0.029  &   13.920     0.051   &    13.372     0.080 \\
24 & 20:34:00.474 &  -32:02:38.43 &    17.707     0.053  &   16.833     0.041  &   16.307     0.035   &    15.873     0.079 \\
25 & 20:33:57.672 &  -32:01:02.46 &    16.610     0.053  &   15.841     0.033  &   15.367     0.051   &    14.979     0.088 \\
26 & 20:34:01.183 &  -32:00:02.73 &    16.535     0.057  &   15.962     0.035  &   15.593     0.046   &    15.264     0.070 \\
27 & 20:34:00.268 &  -31:58:22.97 &    17.099     0.028  &   16.428     0.030  &   16.014     0.051   &    15.648     0.062 \\
28 & 20:34:08.158 &  -31:57:16.48 &    16.710     0.011  &   15.213     0.029  &   14.229     0.049   &    13.177     0.067 \\
29 & 20:33:59.155 &  -31:57:09.43 &    16.886     0.037  &   15.662     0.026  &   14.842     0.022   &    14.274     0.077 \\
30 & 20:34:08.192 &  -31:56:53.97 &    17.899     0.011  &   17.300     0.013  &   16.881     0.010   &    16.549     0.054 \\
31 & 20:34:07.231 &  -31:56:43.23 &    17.737     0.026  &   17.125     0.043  &   16.745     0.036   &    16.382     0.052 \\
32 & 20:34:06.013 &  -31:55:43.34 &    15.450     0.034  &   14.718     0.032  &   14.296     0.043   &    13.918     0.080 \\
33 & 20:34:10.459 &  -31:55:13.60 &    15.872     0.028  &   15.244     0.045  &   14.878     0.044   &    14.508     0.062 \\
34 & 20:34:17.120 &  -31:56:23.58 &    17.916     0.102  &   17.413     0.060  &   17.027     0.051   &    16.678     0.035 
\enddata

\tablecomments{Uncertainties are adjacent to measurements and are at
  the 68\% confidence level.}

\label{tab:localseq}
\end{deluxetable*}

The entire class of SNe IIb and Ibc are collectively known as
``stripped-envelope'' events \citep{Clocchiatti97}. The degree of H
and He envelope deficiency is believed to be the consequence of
varying degrees of a SN progenitor star's mass-loss.  Potential
mechanisms for mass-loss include steady winds \citep{Puls08}, eruptive
mass-loss episodes \citep{Smith06}, or mass transfer due to Roche lobe
overflow in a close binary system \citep{Podsiadlowski92}. It is an
open issue which of these processes dominate, and whether they
occur in relatively high-mass single star progenitors ($M \ga
20\;M_{\odot}$), in lower-mass progenitors ($M \ga 10\;M_{\odot}$),
or a mixture of both.

The wide diversity in the small number of well-observed objects has
complicated attempts at establishing firm connections between SN
subtypes and their progenitor systems. This has been particularly the
case for the inferred progenitors of SNe IIb, which have been quite
varied among a few nearby cases. For example, late-time spectroscopic
and photometric observations of SN\,1993J showed compelling evidence
for a massive binary companion to the red supergiant progenitor star
\citep{Aldering94,Maund04}. However, pre-explosion imaging with the
{\sl Hubble Space Telescope} of SN\,2008ax suggest that it most likely
came from a more compact progenitor, such as a single massive
Wolf-Rayet (WR) star or an interacting binary in a low-mass cluster
\citep{Crockett08}. Most recently, there has been considerable
discourse about a yellow supergiant discovered in pre-explosion
imaging at the location of SN\,2011dh and its viability as a candidate
progenitor
\citep{Maund11,vanDyk11,Arcavi11,Soderberg11,Murphy11,Bietenholz12,Georgy12}.

\begin{figure}[htp!]
\centering
\includegraphics[width=0.95\linewidth]{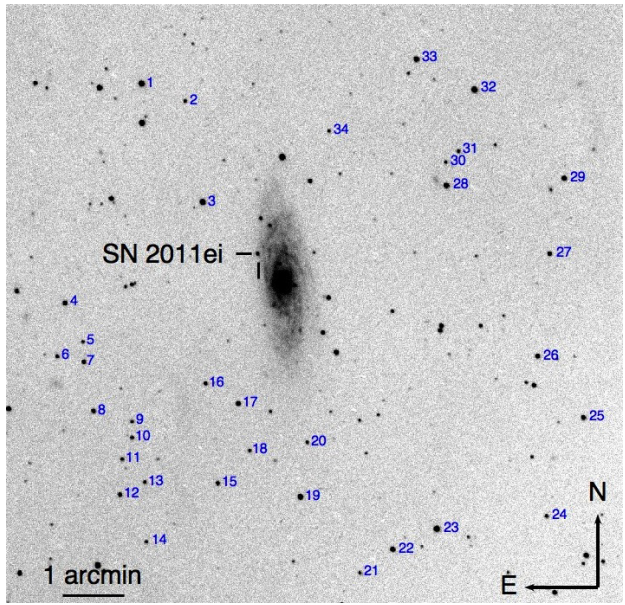}

\caption{PROMPT image of SN\,2011ei around the time of maximum light
  and its host galaxy NGC 6925. Local sequence stars used to calibrate
  stars are numbered according to Table~\ref{tab:localseq}.}

\label{fig:image:11ei}
\end{figure}

From a statistical standpoint, the relatively high frequency of
stripped-envelope SNe ($\sim$30 per cent of all core-collapse events
by volume; \citealt{Li11}) in combination with many non-detections of
progenitor stars in pre-explosion imaging \citep{Smartt09} argues that
at least some progenitor scenarios must involve massive stars in
binary systems
\citep{Podsiadlowski92,Fryer07,Eldridge08,Smith11}. Observational and
theoretical investigations have suggested binary mass-loss scenarios
that could lead to a spectroscopic sequence of SN types from IIb
$\rightarrow$ Ib $\rightarrow$ Ic (e.g.,
\citealt{Filippenko94,Woosley94,Nomoto95,Heger03,Yoon10,Dessart11}).
\citet{Chevalier10} point out that the sequence may only be applicable
for a subset of this population in light of optical and radio data
that favor a division among SNe IIb progenitors between objects like
SN\,1993J that have extended envelopes ($R_* \sim 10^{13}$ cm;
referred to as Type ``eIIb''), and others like SN\,2008ax that are
more compact ($R_* \sim 10^{11}$ cm; Type ``cIIb''). They argue that
Type cIIb closely resemble Type Ib, and that the two subtypes may be
related by a gradual transition depending on remaining H mass.

One key consequence of the proposed stripped-envelope sequence is that
a thin hydrogen layer is expected to be present in many SNe Ib
progenitors at the pre-supernova stage, and it is theorized that it may
be possible to see hydrogen in their early spectra.  Thus, the
detection of hydrogen absorption lines at high velocity in SNe Ib
is important because of its implications for the nature and
evolutionary stage of the progenitor stars.

However, the existence of hydrogen in SNe Ib remains somewhat
uncertain. Reports of suspected high-velocity ($\ga 12,000$ \kms)
hydrogen absorption around 6250 \AA\ exist; e.g., SN\,1999dn,
SN\,2000H \citep{Branch02}; SN\,2005bf
\citep{Anupama05,Folatelli06,Parrent07}, and SN\,2008D
\citep{Soderberg08}. Alternative line identifications, however, have
been offered. For instance, \ion{C}{2} $\lambda$6580, being less than
800 \kms\ to the red of H$\alpha$, has been considered a plausible
identification (e.g., \citealt{Harkness87,Deng00}), and \ion{Si}{2}
$\lambda$6355 and \ion{Ne}{1} $\lambda$6402 have also been suggested
\citep{Hamuy02,Tanaka09,Branch72,Benetti02}.

The uncertain identification of hydrogen in SNe Ib may be a
consequence of current limitations in spectroscopic follow-up of SN
candidates. Because spectra of SNe shortly after outburst are
relatively rare, there is the possibility that SNe IIb vs.\ Ib (vs.\
Ic?)  classification may be biased by the time of observation (see
discussions by \citealt{Branch02} and \citealt{Chornock11}). Indeed,
precise rates of these events may be compromised by the dearth of
early-time spectroscopic and photometric observations several days
prior to maximum light when the SN is faint and the photosphere still
resides in the outer layers of the progenitor star. Almost half of the
ejecta mass may have passed through the photosphere by the time peak
brightness is reached, thus observations within a few days of
explosion offer useful constraints on the progenitor star's current
evolutionary state \citep{Dessart11,Yoon10}.

Our discovery and subsequent monitoring of SN\,2011ei is an example of
the utility of very early data. Discovered in NGC 6925 on July 25.434
UT (all future dates also in UT), early optical spectra of SN\,2011ei
initially showed line features similar to those of the Type IIb
SN\,1993J and SN\,1996cb, but with a noticeably broader H$\alpha$
P-Cygni profile reminiscent of the energetic Type IIb SN\,2003bg
\citep{Stritzinger11,Milisavljevic11}. However, spectra taken
approximately one week later showed rapid evolution away from strong
hydrogen features to conspicuous Type Ib-like helium absorptions.

This rapid change in SN\,2011ei's spectral properties along with a
contemporaneous radio detection motivated a coordinated,
multi-wavelength monitoring program presented here. In
Section~\ref{sec:Observations}, \ref{sec:lightcurves},
\ref{sec:spectra}, and \ref{sec:diagnostics} we present and analyze
X-ray, ultraviolet (UV), optical, and radio observations from
\emph{Swift}, \emph{Chandra}, and ground-based observatories including
the Southern African Large Telescope (SALT), Magellan, and the Karl
G.\ Jansky Very Large Array (VLA). In Section \ref{sec:Discussion} we
discuss the implications these findings have for SN\,2011ei's
progenitor star and its mass-loss history. Finally, in Section
\ref{sec:Conclusions}, we summarize our findings and discuss their
relevance to a wider subset of stripped-envelope core-collapse SNe.


\section{Observations}
\label{sec:Observations}

\subsection{Distance and Reddening}
\label{sec:distance}

Located at coordinates $\alpha = 20^{\rm h}34^{\rm m}22\fs62$ and
$\delta = -31\degr58'23\farcs6$ (J2000.0), SN\,2011ei is situated in
projection along the outer periphery of its host spiral galaxy NGC
6925 (see Figure~\ref{fig:image:11ei}). Estimates of the distance to
NGC 6925 via Tully-Fisher measurements range between 23.3 Mpc and 35.2
Mpc \citep{Willick97}. For all calculations we adopt a distance $D
= 28.5 \pm 5.7$ Mpc and a distance modulus $\mu = 32.27 \pm 0.43$
\citep{Springob09}.
 
The reddening due to the Milky Way in the direction of NGC 6925 is
$E(B-V)_{mw} = 0.059$ mag according to the infrared dust maps of
\citet{Schlegel98}. NGC 6925 is highly inclined which would suggest
possibly significant local extinction along its disk plane. The
supernova, however, appears to lie on the galaxy's near side. We used
the empirically derived relation between equivalent width (EW) of
\ion{Na}{1}\,D absorption and $E(B-V)$ described in \citet{Turatto03}
to estimate the host extinction (though for a discussion of possible
problems with this relationship, see, e.g.,
\citealt{Poznanski11}). The EW(\ion{Na}{1}\,D) = $1.14 \pm 0.10$ \AA\
absorption at the redshift of NGC 6925 determined from the weighted
average of five spectra suggests a host extinction of $E(B-V)_{host} =
0.18$ mag. Combining the inferred host and Galactic extinction, we
adopt a total reddening $E(B-V) = 0.24$ mag and assume $R_V =
A_V/E(B-V) =3.1$. In Section~\ref{sec:lightcurves} we show that this
estimate of the reddening is likely an upper limit.

\subsection{X-ray Observations with Swift-XRT and Chandra} 

We promptly requested X-ray observations from the \emph{Swift}
spacecraft \citep{Gehrels04} using the XRT instrument
\citep{Burrows05} beginning 2011 August 3.2. The total exposure time
was 54~ks.  Data were analyzed using the latest version of the HEASOFT
package available at the time of writing (version 6.11) and corresponding
calibration files. Standard filtering and screening criteria were
applied. No X-ray source is detected coincisdent with SN~2011ei with a
3 sigma upper limit of $4\times10^{-3}\,\rm{c\,s^{-1}}$ in the 0.3-10
keV band. The Galactic neutral hydrogen column density in the
direction of SN\,2011ei is $4.76\times10^{20}\,\rm{cm^{-2}}$
\citep{Kalberla05}.  Assuming a spectral photon index $\Gamma=2$, this
upper limit translates to an absorbed (unabsorbed) flux of
$1.4\times10^{-13}\,\rm{erg\,cm^{-2}\,s^{-1}}$
($1.6\times10^{-13}\,\rm{erg\,cm^{-2}\,s^{-1}}$), corresponding to a
luminosity of $1.5 \times 10^{40}\,\rm{erg\,s^{-1}}$.

From the \emph{Swift}-XRT images, X-ray contamination from the host
galaxy nucleus is clearly identified at the SN site. Consequently, a
10~ks \emph{Chandra} X-ray Observatory observation was also obtained
on 2011 August 21 to resolve SN emission from the host galaxy nucleus
(PI Soderberg, Prop.\ 12500613). Data were reduced with the CIAO
software package (version 4.3) with calibration database CALDB
(version 4.4.2).  Standard filtering using CIAO threads for ACIS data
were applied.

\begin{deluxetable*}{llrccll}
\footnotesize
\centering
\tablecaption{Summary of optical spectroscopic observations of SN\,2011ei}
\tablecolumns{7}
\tablewidth{0pt}
\tablehead{\colhead{Date}                        &
           \colhead{JD}                          &
           \colhead{Phase\tablenotemark{a}}      &
           \colhead{Wavelength Range}            &
           \colhead{Resolution\tablenotemark{b}} &
           \colhead{Telescope/}                  &
           \colhead{Exposure}                    \\
           \colhead{(UT)}                        &
           \colhead{$-2400000$}                  &
           \colhead{(days)}                      &
           \colhead{(\AA)}                       &
           \colhead{(\AA)}                       &      
           \colhead{Instrument}                  &
           \colhead{(s)}                                         
}
\startdata

2011 Jul 29 & 55772.32 & -14 & $3200-9000$  & ~6  & SALT/RSS       & $2 \times 300$ \\
2011 Aug 02 & 55776.33 & -10 & $3200-9000$  & ~6  & SALT/RSS       & $2 \times 300$ \\
2011 Aug 06 & 55780.31 &  -6 & $3200-9000$  & ~6  & SALT/RSS       & $2 \times 600$ \\
2011 Aug 08 & 55782.30 &  -4 & $3400-8800$  & ~6  & SALT/RSS       & $1 \times 600$ \\
2011 Aug 09 & 55783.30 &  -3 & $5900-9000$  & ~6  & SALT/RSS       & $1 \times 600$ \\
2011 Aug 16 & 55789.54 &   3 & $3400-9000$  & ~6  & SALT/RSS       & $2 \times 600$ \\
2011 Aug 21 & 55794.62 &   8 & $3600-8900$  & 13 & SOAR/Goodman   & $3 \times 450$ \\
2011 Aug 26 & 55799.50 &  13 & $3400-9000$  & ~6  & SALT/RSS       & $1 \times 600$ \\
2011 Aug 29 & 55803.48 &  17 & $3400-9000$  & ~6  & SALT/RSS       & $1 \times 600$ \\
2011 Sep 20 & 55824.43 &  38 & $3500-9500$  & ~4  & Magellan/IMACS & $1 \times 900$ \\
2011 Sep 30 & 55834.63 &  48 & $3800-9400$  & ~6  & Magellan/LDSS3 & $3 \times 900$ \\
2011 Oct 18 & 55852.55 &  66 & $3700-9200$  & 18 & NTT/EFOSC2     & $2 \times 1800$ \\
2011 Oct 24 & 55859.34 &  73 & $3400-8800$  & ~6  & SALT/RSS       & $1 \times 600$ \\
2011 Nov 16 & 55881.55 &  95 & $3800-8900$  & 13 & SOAR/Goodman   & $2 \times 2700$ \\
2011 Nov 18 & 55883.55 &  97 & $3500-9500$  & ~4  & Magellan/IMACS & $1\times 1200$ \\
2012 Jun 18 & 56096.33 & 310 & $3500-9600$  & 14 & VLT/FORS2      & $2\times 1800$
\enddata \tablenotetext{a}{Phase is with respect to $V$-band maximum
on JD 2455786.5 (2011 Aug 13.0).}  \tablenotetext{b}{FWHM of night
sky emission lines.}
\label{tab:specobservations}
\end{deluxetable*}

\begin{figure*}[htp!]
\centering
\includegraphics[width=0.70\linewidth]{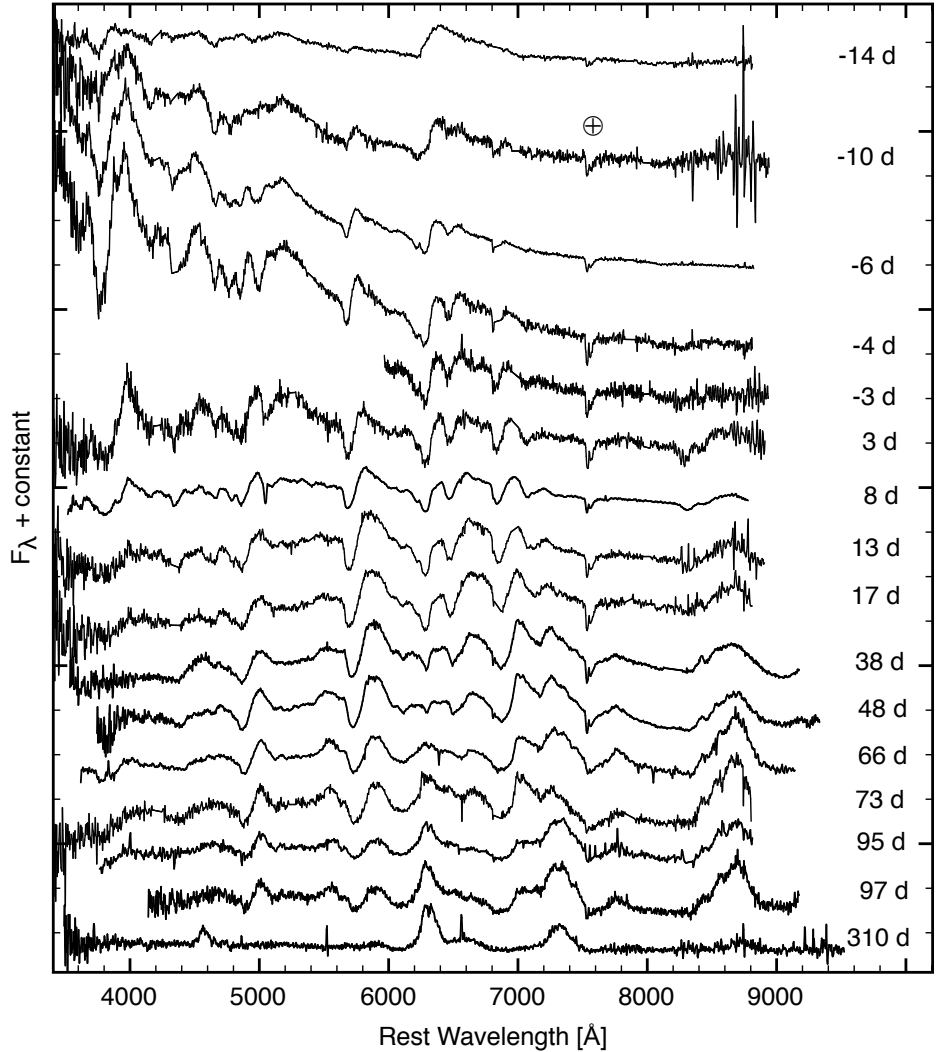}

\caption{Optical spectra of SN\,2011ei. Phase is with respect to
  $V$-band maximum. Telluric features have been
  marked with the ``$\oplus$'' symbol.}

\label{fig:spectra:11ei}
\end{figure*}

Consistent with the \emph{Swift}-XRT observation, no X-ray source is
detected at the SN position in the \emph{Chandra} observation with a
3$\sigma$ upper limit of $7.6\times 10^{-4}\,\rm{c\,s^{-1}}$ in the
0.5-8 keV band. From the count-rate, we derived an absorbed flux limit
of $5.5\times10^{-15}\;\rm{erg\,s^{-1}\,cm^{-2}}$. Using the Galactic
relations between $\rm N_H$ and $A_V$ \citep{Predehl95,Watson11}, an
intrinsic hydrogen column density $\rm N_{H[int]} \sim 1.5 \times
10^{21}\; \rm{cm^{-2}}$ is inferred. Accounting for the total
absorption (Galactic plus intrinsic) leads to a unabsorbed flux limit
of $7\times10^{-15}\;\rm{erg\,s^{-1}\,cm^{-2}}$ (assuming a simple
power-law model with spectral photon index $\Gamma=2$), corresponding
to an X-ray luminosity of
$L_{X[0.5-8keV]}<7\times10^{38}\,\rm{erg\,s^{-1}}$. By comparison,
X-ray emission at similar epochs from a SN IIb such as SN\,1993J would
have been detected ($L_{X[0.3-8keV]} \approx 8 \times 10^{39}$
erg~s$^{-1}$), whereas the emission from an object like SN\,2011dh
($L_{X[0.3-8keV]} \approx 1.5 \times 10^{38}$ erg~s$^{-1}$) would not
have been detected \citep{Soderberg11}.

\subsection{Ultraviolet and Optical Photometry}

UV/optical observations of SN\,2011ei with the \emph{Swift}-UVOT
instrument \citep{Roming05} began 2011 August 3.2 and continued
through to 2011 September 10.9.  Data were acquired using all of its
six broad band filters that span the wavelength range of approximately
$1600-6000$ \AA. The \emph{Swift}-UVOT photometry presented in
Table~\ref{tab:phot} is based on the photometric system described in
\citet{Poole08}.  In this system the \emph{Swift} $b$ and $v$ filters
are roughly equivalent to the standard Johnson Kron-Cousins $B$ and
$V$ filters (see \citealt{Poole08} for details). 

We analyzed the \emph{Swift}-UVOT photometric data following the
prescriptions of \cite{Brown09}. A $3''$ aperture was used to maximize
the signal-to-noise (S/N) ratio. Late-time images acquired 2012 March
27 -- April 6 have been used to estimate and remove the underlying
host galaxy contribution. Observations beyond 2011 September 08 are
not reported because supernova emission was too faint to obtain
reliable measurements. Also not reported are observations in the
$uvw2$ and $uvm2$ filters where the SN was not detected with
significant S/N.

Also reported in Table~\ref{tab:phot} is {\it BVRI} photometry acquired with
the Panchromatic Robotic Optical Monitoring and Polarimetry Telescope
(PROMPT; \citealt{Reichart05}) PROMPT3 and PROMPT5 located at Cerro
Tololo Inter-American Observatory. All images were dark subtracted and
flat-field corrected. Frames taken with the same filter were stacked
in order to produce a final deeper image.  The instrumental magnitude
of the SN was measured in all the stacked frames using the PSF
technique after a scaled template image was subtracted from the SN
image. SN photometry was calibrated against a sequence of stars
located close to the SN. This local sequence, with magnitudes reported
in Table~\ref{tab:localseq}, was calibrated to the standard Johnson
Kron-Cousins photometric systems using observations of
\citet{Landolt92,Landolt07} standard stars.

In addition to \emph{Swift}-UVOT and PROMPT photometry, we estimate an
apparent magnitude $R = 18.0 \pm 0.4$ on 2011 July 25.434 (JD
2455767.93) based on unfiltered discovery images of the SN. The data
were obtained with the 0.35-m Celestron telescope equipped with a
KAF-3200ME camera most sensitive to $R$-band emission at Parkdale
Observatory, Canterbury, New Zealand. After flat-fielding and
correcting for the bias and dark current levels, SN emission was
calibrated against field comparison stars of the USNO-B 1.0 star
catalog and then converted to the standard $R$ value.

\begin{deluxetable}{lrrc}
\tablecaption{VLA observations of SN\,2011ei}
\tablecolumns{4}
\tablewidth{0pt}
\tablehead{
\colhead{Date} & \colhead{$\nu$} & \colhead{$F_{\nu}$} &
\colhead{Array} \\
\colhead{(UT)} & \colhead{(GHz)} & \colhead{($\mu$Jy)} & \colhead{Config.}}
\startdata
2011 Aug 3.2 & 5.0 & $93\pm 25$ &  A\\
\nodata & 25 & $130\pm 140$ & \nodata \\
2011 Aug 5.2 & 5.0 & $143\pm 30$ & A \\
2011 Aug 7.2 & 3.1 & $37\pm 45$ & A \\
\nodata & 5.0 & $226\pm 38$ & \nodata  \\
\nodata & 8.5 & $589\pm 38$ & \nodata \\
2011 Aug 12.2 & 3.1 & $143\pm 45$ & A \\
\nodata & 5.0 & $289\pm 33$ & \nodata \\
\nodata & 8.5 & $589\pm 38$ & \nodata \\
2011 Aug 17.3 & 3.1 & $289\pm 68$ & A \\
\nodata & 5.0 & $564\pm 33$ & \nodata \\
\nodata & 8.5 & $743\pm 38$ & \nodata \\
2011 Aug 25.2 & 3.1 & $353\pm 52$ & A \\
\nodata & 5.0 & $511\pm 74$ &\nodata \\
2011 Sep 8.1 & 3.1 & $497\pm 65$ & A \\
\nodata & 5.0 & $279\pm 50$ & \nodata \\
\nodata & 8.5 & $182\pm 47$ & \nodata \\
2011 Oct 11.0 & 5.0 & $411\pm 22$ & D \\
\nodata & 6.8 & $332\pm 19$ & \nodata \\
\nodata & 8.4 & $290\pm 26$ & \nodata \\
2011 Dec 3.9 & 5.0 & $256\pm 30$ & D \\
\nodata & 6.8 & $168\pm 19$ & \nodata \\
\nodata & 8.4 & $219\pm 19$ & \nodata \\
2012 Jan 29.8 & 5.0 & $272\pm 26$ & C \\
\nodata & 6.8 & $197\pm 18$ & \nodata \\
\nodata & 8.4 & $124\pm 20$ & \nodata \\
2012 Mar 10.6 & 2.5 & $268\pm 41$ & C \\
\nodata & 3.5 & $330\pm 30$ & \nodata \\
\nodata & 5.0 & $199\pm 20$ & \nodata \\
\nodata & 6.8 & $116\pm 15$ & \nodata \\
\nodata & 8.4 & $130\pm 16$ & \nodata 
\enddata
\label{tab:vla}
\end{deluxetable}

\subsection{Optical Spectroscopy}

A total of 16 epochs of long slit optical spectra of SN\,2011ei were
collected. The details of all observations are provided in
Table~\ref{tab:specobservations} and the reduced spectra are plotted
in Figure~\ref{fig:spectra:11ei}. Most observations were made at SAAO
Observatory with the 10-m Southern African Large Telescope (SALT)
using the Robert Stobie Spectrograph (RSS; \citealt{Burgh03}). The
holographic grism pg0900 was used at combinations of four tilts to
span a wavelength window of $3200-9000$ \AA\ and cover gaps between
CCD detectors. On chip CCD $4 \times 4$ binning in the spectral and
spatial directions was employed, yielding a dispersion of 1.96 \AA\
pixel$^{-1}$ and $\approx 6$ \AA\ full-width-at-half-maximum (FWHM)
resolution. An atmospheric dispersion corrector minimized the effects
of atmospheric refraction for all observations which were made
generally around an airmass of 1.2. Conditions were mostly
non-photometric with a median seeing around $1\farcs3$.

Additional spectra were obtained using a variety of telescopes and
instrumental setups. The Magellan 6.5-m Baade and Clay telescopes
located Las Campanas Observatory were used with the IMACS
\citep{Bigelow98} and
LDSS-3\footnote{http://www.lco.cl/telescopes-information/magellan/instruments-1/ldss-3-1/}
instruments, respectively. The 3.6-m New Technology Telescope (NTT)
located at La Silla was used with the
EFOSC2\footnote{http://www.eso.org/sci/facilities/lasilla/instruments/efosc/}
instrument, and the Southern Astrophysical Research (SOAR) 4.1-m
telescope located at Cerro Pach\'on was used with the Goodman
spectrograph \citep{Clemens04}. Additional late-time spectra were
obtained with the Very Large Telescope (VLT) with the FOcal Reducer
and low dispersion Spectrograph
(FORS2\footnote{http://www.eso.org/sci/facilities/paranal/instruments/fors/}).

Reduction of all spectra followed standard procedures using the
IRAF/PyRAF software. SALT data were first processed by the
PySALT\footnote{http://pysalt.salt.ac.za/} pipeline (see
\citealt{Crawford10}). Wavelength calibrations were made with arc
lamps and verified with the night sky lines. Relative flux
calibrations were made from observations of spectrophotometric
standard stars from \citet{Oke90} and \citet{Hamuy92,Hamuy94}. Gaps
between CCD chips have been interpolated in instances when dithering
between exposures was not possible, and cosmetic defects have been
cleaned manually.

A recession velocity of 2780 \kms\ determined from coincident nebular
H$\alpha$ emission has been removed from the presented spectra. This
velocity is in agreement with a previous measurement of $2792 \pm 4$
\kms \ reported by \citet{Koribalski04} using radio \ion{H}{1} lines
from the host NGC 6925.

\subsection{Radio Observations with the VLA}

We also obtained radio observations of SN\,2011ei beginning shortly
after explosion.  On 2011 August 3.2, we observed SN\,2011ei with the
Karl G.\ Jansky Very Large Array (VLA; \citealt{Perley11}) at
frequencies of $\nu=5$ and 25 GHz. Following a strong detection of
radio emission at 5 GHz \citep{Chomiuk11}, we initiated a VLA
monitoring campaign at frequencies spanning 2-9 GHz to obtain
observations between August 2011 - March 2012 (PI Soderberg; 10C-168,
11B-212).

All VLA observations were obtained with the widest bandwidth available
at any time during the commissioning phase.  Prior to 2011 October 1
this amounted to two 128-MHz wide sub-bands (total bandwidth of 256
MHz), while we later utilized the total bandwidth of 2 GHz at S- and
C-bands; in the latter case the central frequencies were tuned to 5.0
and 6.8 GHz.  Given that fully upgraded X-band receivers were not yet
available, our observations at this band spanned 8.0-8.8 GHz.  We used
J2101-2933 as a phase-calibrator source and 3C48 for flux calibration.
Initial observations were obtained in the extended A-configuration
while later observations were obtained in the compact C- and
D-configurations.

Data were reduced using the standard packages of the Astronomical
Image Processing System (AIPS).  In each observation, we fit a
Gaussian to the source and extracted the integrated flux density. For
observations in the compact D-array we limited the uv-range ($\gtrsim
5~\rm k\lambda$) to minimize contamination from diffuse host galaxy
emission.  Our final flux density measurements for SN\,2011ei are
reported in Table~\ref{tab:vla}.

\begin{figure}[htp!]
\centering
\includegraphics[width=\linewidth]{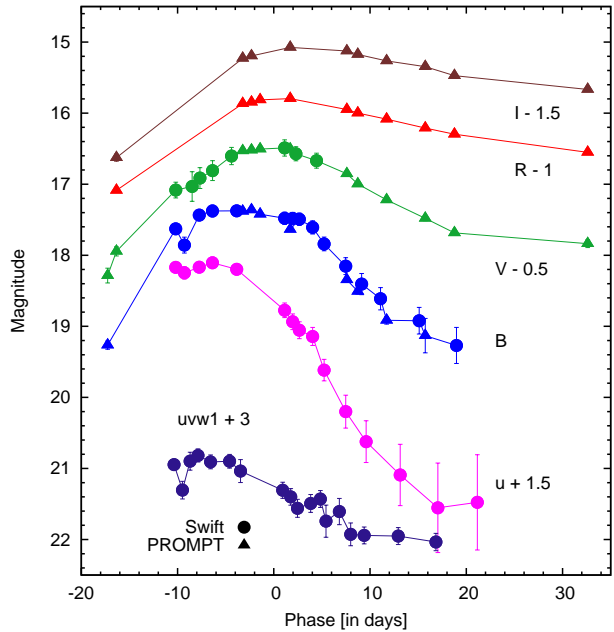}

\caption{\emph{Swift}-UVOT and PROMPT light curves of
  SN\,2011ei. Light curves have been shifted by the indicated amounts for
  clarity.}

\label{fig:lightcurves}
\end{figure}

\begin{deluxetable}{cccl}[htp!]
\footnotesize
\centering
\tablecaption{Epoch of maximum light and peak
  magnitude for the light curves of SN\,2011ei}
\tablecolumns{4}
\tablewidth{0pt}
\tablehead{\colhead{Filter}                       &
           \colhead{Peak Time}                    &
           \colhead{Peak Time}                    &
           \colhead{Peak Obs.}                    \\
           \colhead{}                             &
           \colhead{(UT)}                         &
           \colhead{(JD$-2400000$)}               &
           \colhead{(m($\lambda$)$_{\rm max}$)}}
\startdata
$uvw1$ & 2011 Aug 05.0 & 55778.5 $\pm 2.0$  & $17.8 \pm 0.10$ \\ 
$u$    & 2011 Aug 06.0 & 55779.5 $\pm 1.5$  & $16.6 \pm 0.10$ \\
$B$    & 2011 Aug 10.0 & 55783.5 $\pm 2.0$  & $17.4 \pm 0.05$ \\
$V$    & 2011 Aug 13.0 & 55786.5 $\pm 1.5$  & $17.0 \pm 0.05$ \\
$R$    & 2011 Aug 13.5 & 55787.0 $\pm 1.5$  & $16.8 \pm 0.07$ \\ 
$I$    & 2011 Aug 16.0 & 55789.5 $\pm 1.5$  & $16.6 \pm 0.07$
\enddata
\label{tab:lightcurve}
\end{deluxetable}

\begin{figure}[htp!]
\centering
\includegraphics[width=\linewidth]{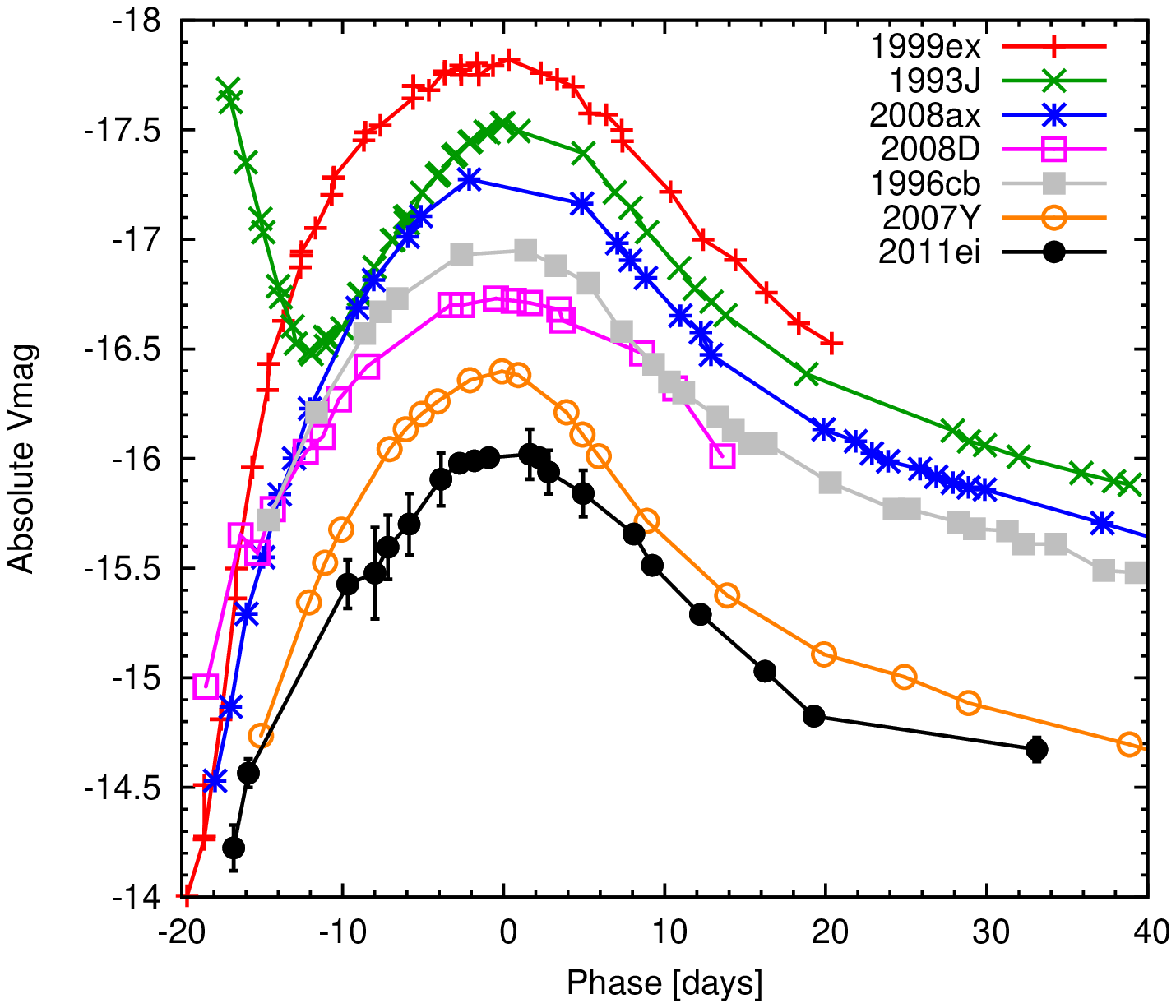}

\caption{$V$-band absolute light curve of SN\,2011ei compared to light
  curves of Type IIb and Ib objects: SN\,1999ex (Type Ib;
  \citealt{Stritzinger02}), SN\,1993J (Type IIb; \citealt{Lewis94}),
  SN\,2008ax (Type IIb; \citealt{Pastorello08}), SN\,1996cb (Type IIb;
  \citealt{Qiu99}), and SN\,2007Y (Type Ib; \citealt{Stritzinger09}).}

\label{fig:abs_Vmag}
\end{figure}

\begin{figure*}[htp!]
\centering
\includegraphics[width=0.46\linewidth]{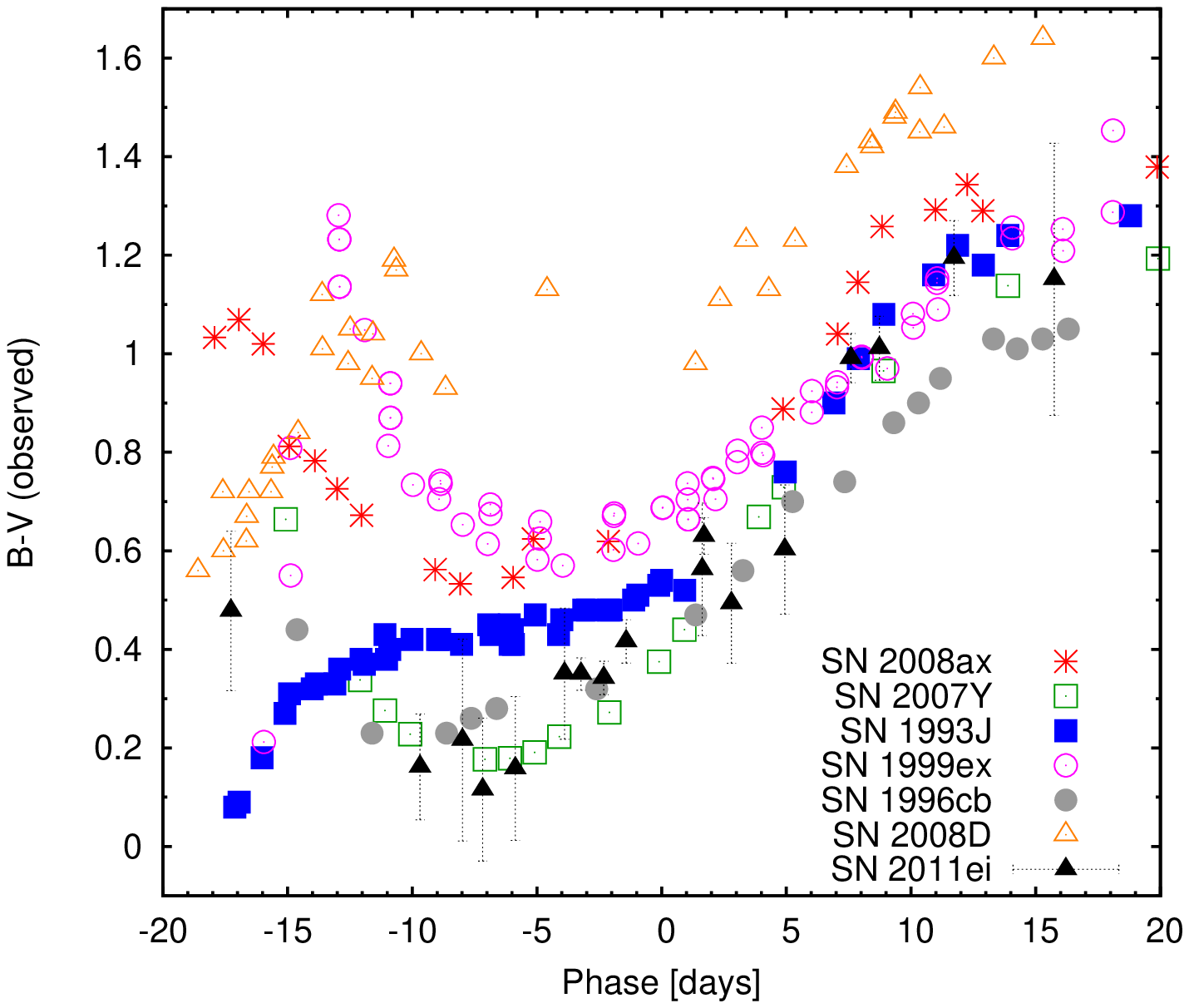}
\includegraphics[width=0.46\linewidth]{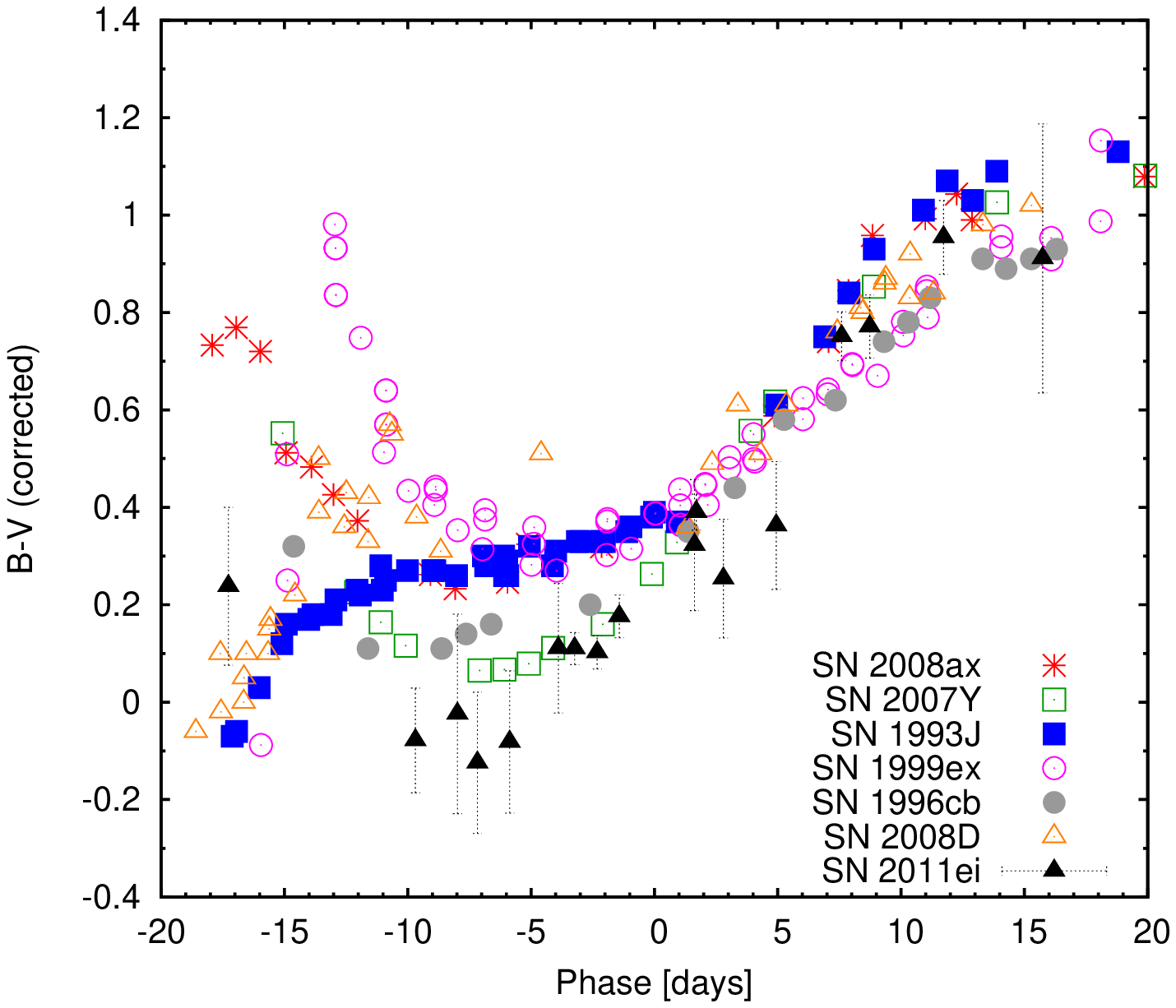}

\caption{The $B-V$ color curves of SN\,2011ei compared to those of other
  stripped-envelope SNe: SN 2008ax (Type IIb;
  \citealt{Pastorello08}), SN 2007Y (Type Ib;
  \citealt{Stritzinger09}), SN 1993J (Type IIb; \citealt{Lewis94}), SN
  1999ex (Type Ib; \citealt{Stritzinger02}), SN 1996cb (Type IIb;
  \citealt{Qiu99}), and SN 2008D (Type Ib; \citealt{Mazzali08}). On the
  left are observed colors, and on the right are colors corrected for
  extinction.}

\label{fig:colorevol}
\end{figure*}


\section{Properties of the UV, Optical,
            and Bolometric Light Curves}

\label{sec:lightcurves}


\subsection{UV and Optical Light Curves and Colors}

Figure~\ref{fig:lightcurves} shows the observed \emph{Swift}-UVOT and
PROMPT light curves of SN\,2011ei. The properties of these curves are
presented in Table~\ref{tab:lightcurve}.  The time and observed
magnitude at maximum light of each filter light curve were estimated
with low-order polynomials.  The average time to rise to peak
brightness from the time of explosion ($t_{\rm exp}$) for SNe IIb and
Ib in the $V$-band is $\Delta t_{\rm exp} \la 20$ d
\citep{Richardson06,Drout11}. Thus, our discovery observation is
likely to have been obtained within $\sim 1$ day of outburst and we
adopt an explosion date of July 25.0 hereafter.

In Figure~\ref{fig:abs_Vmag}, the absolute $V$-band light curve of
SN\,2011ei is compared with those of well observed SNe IIb and
Ib. With a peak absolute magnitude of $M_V \approx -16$ mag,
SN\,2011ei was approximately 2 mag below the mean peak absolute
magnitudes of the entire class of stripped-envelope events
\citep{Richardson06,Drout11}.  Other low-luminosity events
include the Type Ib SN 2007Y ($M_V = -16.5$; \citealt{Stritzinger09}),
and SN 2008D ($M_V = -16.7$; \citealt{Soderberg08}).

In Figure~\ref{fig:colorevol}, we plot the observed $(B-V)$ color
curve of SN 2011ei with respect to $V$-band maximum along with those
of other SNe IIb and Ibc.  Also shown are the same color curves
corrected for extinction. Like SN\,1999ex and SN\,2008ax, the $(B-V)$
colors of SN\,2011ei first become increasingly blue in the days
immediately following outburst. Then, around 5-10 days before maximum,
the SN becomes redder with time and the color indices increase
monotonically.  The majority of $(B-V)$ color indices of SN\,2011ei
are somewhat bluer than those of other SNe after correction. This blue
excess suggests that the adopted $E(B-V) = 0.24$ mag is likely an
upper limit.

\subsection{Bolometric Light Curve}
\label{sec:lbol}

The extinction-corrected \emph{Swift}-UVOT and PROMPT light curves
were combined to obtain a quasi-bolometric light curve
($L_{\rm{bol}}^{\rm{quasi}}$) of SN~2011ei.  Low-order polynomials
have been used to interpolate the light curves. The total UV+{\it
BVRI} flux was determined by summing the integrated fluxes of the
different filters and uncertainties have been propagated following
standard practice.

The quasi-bolometric UV+{\it BVRI} light curve was then transformed
into a bolometric light curve assuming $L_{\rm{bol}}^{\rm{quasi}}
\approx 0.8L_{\rm{bol}}$ and that $0.2L_{\rm{bol}}$ is emitted as
unobserved near infrared emission. This estimate is in line with
bolometric reconstructions of SNe IIb and Ib with extensive
observations such as SN\,2007Y \citep{Stritzinger09}, SN 2008ax
\citep{Pastorello08,Taubenberger11}, and SN\,2008D
\citep{Mazzali08,Soderberg08}.

\begin{figure}[htp!]
\centering
\includegraphics[width=\linewidth]{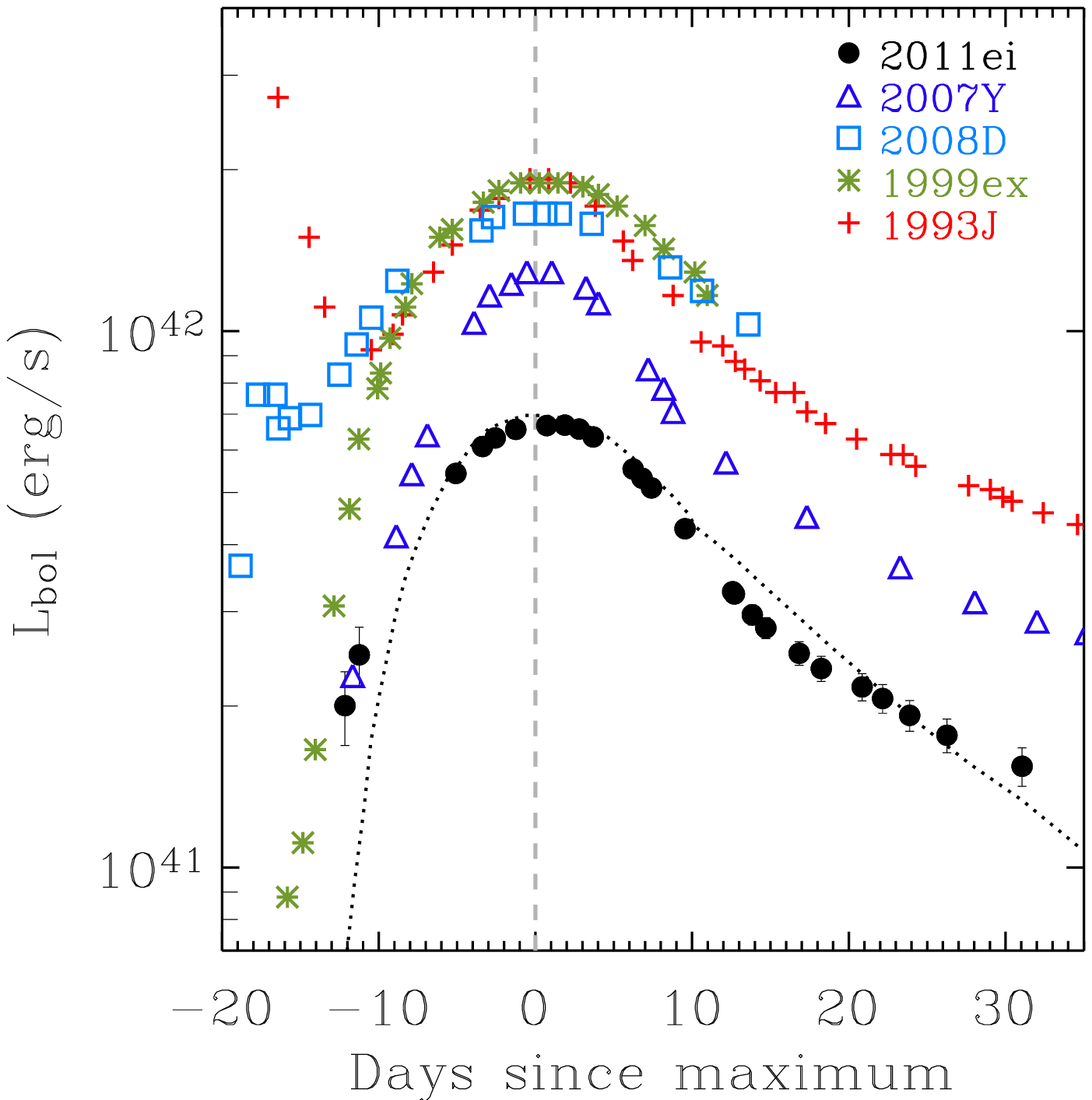}

\caption{Reconstructed bolometric light curve of SN\,2011ei compared
  to those of SN 2007Y (Type Ib; \citealt{Stritzinger09}), SN 2008D
  (Type Ib; \citealt{Mazzali08}), SN\,1999ex (Type Ic;
  \citealt{Stritzinger02}), and SN 1993J (Type IIb;
  \citealt{Richmond94}). Plotted over the points for SN\,2011ei is the
  final best-fit curve used to determine explosion parameters. Phase
  is with respect to the peak of SN\,2011ei's bolometric light curve,
  which occurs on 2011 August 08.4 (JD 2456147.9).}

\label{fig:lbol}
\end{figure}

The final bolometric light curve of SN 2011ei is shown in
Figure~\ref{fig:lbol}. Also shown in the figure are the bolometric
light curves of SN\,1999ex, 1993J, 2007Y, and 2008D . The bolometric
light curve peaks $\approx14.4$ days after outburst with $L_{\rm bol} \sim
7\times 10^{41}\,\rm{erg\,s^{-1}}$, making SN\,2011ei one of the least
luminous SNe IIb/Ib observed to date. Among the closest in luminosity
is SN 2007Y which peaked at $L_{\rm bol} \sim 1.3 \times 10^{42}\;{\rm erg
s}^{-1}$.

\subsection{Explosion Parameters}
\label{sec:explosion}

We modeled of the bolometric light curve to infer the ejecta mass
($M_{ej}$), the nickel mass ($M_{Ni}$) and the kinetic energy of the
ejecta ($E_{k}$). Following the procedures of \cite{Valenti08a}, we
assumed the early-time ($\Delta t_{\rm exp} < 40$ d) light curve to
correspond with the photospheric regime and adopted a constant optical
opacity $k_{\rm{opt}}=0.07\,\rm{cm^{2}\,g^{-1}}$, which is valid so
long as electron scattering is the dominant opacity source and the
ejecta are triply ionized (see, e.g., \citealt{Chugai00}). At late
times ($\Delta t_{\rm exp}>40$ d) the optical depth of the ejecta
decreases and the observed luminosity is powered by $\gamma$-rays
arising from the $^{56}$Co decay, $\gamma$-rays from electron-positron
annihilation, and the kinetic energy of the positrons
(\citealt{Sutherland84,Cappellaro97}).

From the above modeling the following best-fitting parameters were
derived: $M_{ej} \sim 1.6\;\rm{M_{\sun}}$, $E_{k} \sim
2.5\times10^{51}\;\rm{erg}$, and $M_{Ni} = 0.030 \pm
0.010\;\rm{M_{\sun}}$.  Our model under-predicts the observed
luminosity at early times ($\Delta t_{\rm exp} < 5$ d). This suggests
the presence of an additional component of emission. We explore the
possibility of cooling envelope emission following shock break-out in
Section \ref{sec:radius}.

As an independent check we note that our best-fitting explosion
parameters imply a photospheric velocity, $v_{\rm{ph}}$, at peak
luminosity of
\begin{eqnarray}
v_{\rm{ph}}\approx(\frac{6E_{k}}{5 M_{ej}})^{0.5}\sim
10,000\,\rm{km\,s^{-1}}, \nonumber
\end{eqnarray}
which is in agreement with the expansion velocities observed in the
optical spectra (see Section \ref{sec:earlyspectra}). A separate check
of the Ni mass comes from a relationship between $M_{Ni}$ vs.\ peak
absolute $R$-band magnitude ($M_{R}$) presented in
\citet{Drout11}. Given SN\,2011ei's extinction-corrected peak of
$M_{R} \approx -16.1$~mag, this relationship yields a Ni mass of
$M_{Ni} < 0.04\;M_{\sun}$, which is consistent with our result.

\subsection{Limits on the Radius of the Progenitor Star}
\label{sec:radius}

The fortuitous discovery of SN\,2011ei not long after outburst enabled
us to trace the temporal evolution of the early optical emission (see
Figure~\ref{fig:lightcurves}).  As discussed by \citet{Ensman92}, this
is the time of cooling envelope emission after shock breakout, and
provides a good indicator of the initial stellar extent.  

A significant difference between SN\,2011ei and SN 1993J clearly
illustrated in Figure~\ref{fig:abs_Vmag} is that SN\,2011ei does not
show the prominent early-time peak exhibited by the light curve of
SN\,1993J. This peak is attributed to the initial shock heating and
subsequent cooling of a low-mass envelope of SN\,1993J's progenitor
star of radius $\sim 4 \times 10^{13}$ cm
\citep{Bartunov94,Shigeyama94,Woosley94}.  The absence of a comparable
early luminous phase in SN\,2011ei suggests its progenitor star had a
smaller envelope than that of SN\,1993J.

The first $R$-band detection of SN\,2011ei was made shortly after
explosion with luminosity, $L_R\approx 2.4\times 10^{40}~\rm
erg\,s^{-1}$.  This initial observation was used to constrain the
radius of the progenitor star by modeling the cooling envelope
emission adopting the formalism of \citet{Chevalier08}. In this model
the emission is roughly approximated as a blackbody spectrum with a
radius and temperature determined by the ejecta parameters, $M_{\rm
ej}$ and $E_k$ (see \citealt{Chevalier08} and \citealt{Nakar10} for a
detailed discussion).  In turn, these parameters enable an estimate of
the progenitor radius, $R_*$. In this scenario, the temperature of the
photosphere is
\begin{eqnarray}
T_{\rm ph}\approx
7800~E_{k,51}^{0.03}~M_{\rm ej,\odot}^{-0.04} R_{*,11}^{0.25} \Delta
t_{\rm exp}^{-0.48}~\rm K,
\end{eqnarray}
and the photospheric radius is
\begin{eqnarray}
R_{\rm ph}\approx
3\times 10^{14}~E_{k,51}^{0.39}~M_{\rm ej,\odot}^{-0.28} \Delta
t_{\rm exp}^{0.78}~\rm cm.
\end{eqnarray}
Here, $E_{k,51}$ is $E_k$ in units of $10^{51}$ erg, $M_{\rm
ej,\odot}$ is in units of $M_{\odot}$, $R_{*,11}$ is in units of
$10^{11}$ cm, and $\Delta t_{\rm exp}$ is time in days since
explosion.

For explosion parameters determined in Section~\ref{sec:explosion} and
the adopted explosion date 2011 July 25.0, the extinction-corrected
discovery magnitude constrains $R_* < 10^{11}~$ cm.  This estimate is
consistent with upper limits on the progenitor radius derived from a
similar analysis of the first $B$- and $V$-band detections $\sim 2.5$
days after explosion (see Table~\ref{tab:phot}). We note that the
temperature is sufficiently low that recombination is important and
the model of \citet{Chevalier08} is inaccurate. Furthermore, the model
predicts a temperature that decreases with time, whereas our earliest
photometric and spectroscopic measurements suggest an increasing
temperature. However, our estimate of the upper limit to progenitor
radius should still be valid to within a factor of 2--3. Thus, the
early optical emission points to a compact progenitor, similar to
those of SNe Ibc and compact IIb of radius $R_* \lesssim 10^{11}~$cm
\citep{Chevalier10}.

\begin{figure}[htp!]
\centering
\includegraphics[width=\linewidth]{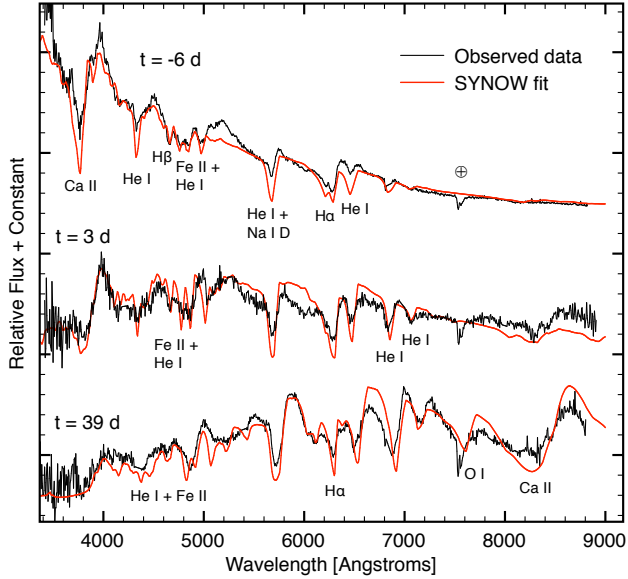}

\caption{Representative SYNOW fits to optical spectra of SN
2011ei. Phase is with respect to $V$-band maximum.}

\label{fig:synowfits}
\end{figure}

\begin{figure}
\centering
\includegraphics[width=\linewidth]{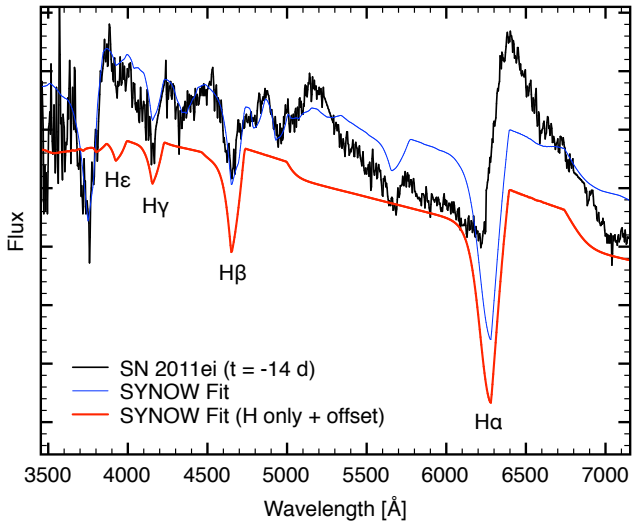}

\caption{Attempted SYNOW fits of $t=-14$~d spectrum of SN
  2011ei. The blue line represents the complete model and the red line
  represents the model only using hydrogen. Features blueward of 5000
  \AA\ are fit reasonably well, but none of the models we tried fit
  the broad H$\alpha$ profile.}

\label{fig:spectra:11ei_honly}
\end{figure}


\section{Optical Spectroscopy}
\label{sec:spectra}


We now describe the optical spectra of SN\,2011ei and compare their
properties to those of other stripped-envelope events. We also use
these spectra to probe SN\,2011ei's ejecta mass distribution by
modeling the SN's late-time emissions. The times of the spectroscopic
observations, $t$, are provided relative to the time of $V$-band
maximum. Line identifications and estimates of expansion velocities
were made with the supernova spectrum synthesis code SYNOW. Manual and
automated procedures employing the recently updated versions of the
software SYN++ in combination with SYNAPPS were
used.\footnote{Software was retrieved from https://c3.lbl.gov/es/} The
basic assumptions of SYNOW include spherical symmetry, velocity
proportional to radius, a sharp photosphere, line formation by
resonant scattering treated in the Sobolev approximation, local
thermodynamic equilibrium (LTE) for the level populations, no
continuum absorption, pure resonance scattering, and only thermal
excitations. Fits are constrained by how we are able to best match
absorption minimum profiles, as well as the relative strengths of all
the features (see \citealt{Branch02} for more description of fitting
parameters and \citealt{Thomas11} for software
details). Representative SYNOW fits are shown in
Figure~\ref{fig:synowfits}.

\subsection{Early Spectral Evolution}
\label{sec:earlyspectra}

The pre-maximum light optical spectra of SN\,2011ei evolved
rapidly. The earliest spectrum ($t=-14$~d;
Figure~\ref{fig:spectra:11ei_honly}) shows strong Type II-like
hydrogen P-Cygni features.  The H$\alpha$ profile exhibits a
`saw-toothed' shape, rising steeply from a minimum at 6220~\AA\
($-16,000$ \kms) to a maximum at 6395 \AA. Redshifted emission extends
to approximately 7000~\AA\ ($+19,000$ \kms). In
Figure~\ref{fig:spectra:11ei_honly}, two SYNOW synthetic fits
to the $t=-14$~d spectrum of SN\,2011ei are shown: one a complete
model, and the other using only hydrogen. The H$\beta$, H$\gamma$, and
H$\epsilon$ absorptions are fit reasonably well with a velocity of
14,500 \kms. Additional features in this spectrum are fit with
\ion{Ca}{2}, \ion{Na}{1}, and \ion{Fe}{2}. Attempts to produce
satisfactory fits of the H$\alpha$ profile failed, likely due to the
LTE and resonance scattering assumptions of SYNOW.

\begin{figure}[htp!]
\centering
\includegraphics[width=0.91\linewidth]{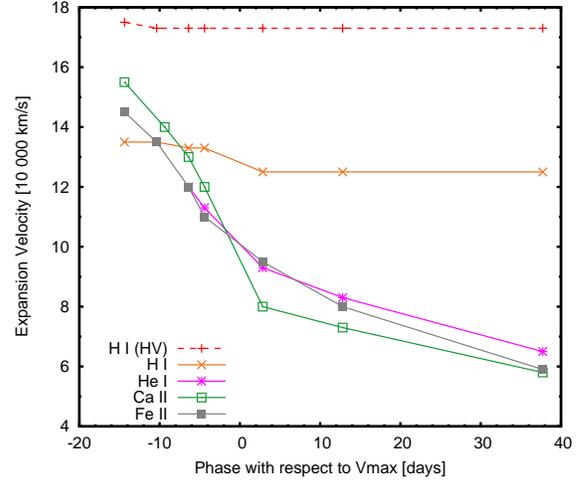}

\caption{Expansion velocities of prominent ions in the spectra of SN
  2011ei determined from SYNOW fitting. Uncertainties in
  velocity estimates are $\sim 500$ \kms. The identification of
  high-velocity (HV) \ion{H}{1} is poorly constrained since it is
  observed only in H$\alpha$.}

\label{fig:velocities}
\end{figure}

In the next two spectra on $t=-10$ and $t=-6$~d, the prominent
H$\alpha$ emission profile subsides and the spectra instead exhibit
conspicuous He absorptions usually associated with SNe Ib. The
continuum becomes increasingly blue and the \ion{He}{1} $\lambda$5016,
$\lambda$5876, $\lambda$6678, and $\lambda$7065 lines appear but are
weak and narrow. Blueshifted absorption of the H$\alpha$ line around
6250~\AA\ develops two minima. The center of the trough sits around
the observed wavelength of 6300~\AA\ so contamination from the
[\ion{O}{1}] $\lambda$6300 night sky line is possible. However, the
peak between the two absorptions is fairly strong and too broad to be
associated with narrow, unresolved telluric emission.

The 6250~\AA\ feature was modeled using two components of hydrogen at
velocities of $\approx$17,000 \kms\ and 14,000 \kms.  Though two
distinct absorptions are observed around H$\alpha$, only one
conspicuous absorption is observed around H$\beta$.  The
SYNOW fit confirms that two components of hydrogen with a
velocity difference of $\sim3000$ \kms\ blend together and manifest as
a single feature in all Blamer lines except H$\alpha$.  Attempts to
fit with other ions such as \ion{Si}{2}, \ion{Ne}{1}, and \ion{C}{2}
introduced inappropriate features elsewhere in the synthetic spectra
and/or were at velocities well below the photosphere and were
considered unsuccessful. The properties of this absorption as observed
in SN\,2011ei and a number of stripped-envelope events is explored
further in Section~\ref{sec:dabsorption}.

From $t = -6$ to $t = +3\rm\;d$, the \ion{He}{1} lines
strengthen. High-velocity calcium absorption is not detected in the
early spectra, which is interesting given that it has been reported at
these epochs in several SNe IIb and Ib including SN\,1999ex,
SN\,2005bf, SN\,2007Y, and SN\,2008ax
\citep{Stritzinger02,Folatelli06,Stritzinger09}, but not in SN\,2008D
\citep{Modjaz09}. Emission around 8700 \AA\ associated with the
infrared \ion{Ca}{2} triplet is noticeable shortly after maximum light
and then gradually strengthens in the following months.

Additional ions outside of \ion{H}{1}, \ion{He}{1}, \ion{Ca}{2},
\ion{Na}{1}, and \ion{Fe}{2} were introduced to the synthetic spectra
to model the $t = 39$ d spectrum. The most conspicuous change is
\ion{O}{1} absorption around 7600 \AA. Additional absorptions are fit
using a blend of \ion{S}{2} features around 5200~\AA, \ion{Si}{2}
$\lambda$6355, and \ion{Ba}{2} $\lambda$6142 and $\lambda$6496,
however, their removal do not change any conclusions reported here and
they are not considered significant.

The evolution in the expansion velocities for the most prominent ions
in SN\,2011ei as estimated from the SYNOW fitting is shown in
Figure~\ref{fig:velocities}. Some noteworthy trends are observed. The
high-velocity hydrogen shows almost no change over all of the sampled
epochs, and the second, lower-velocity hydrogen component shows a
slight drop before leveling out after maximum light. The remaining
ions \ion{He}{1}, \ion{Ca}{2}, and \ion{Fe}{2} show velocities that
drop most steeply before maximum light, after which the rate of
velocity change decreases. The expansion velocities of the \ion{He}{1}
and \ion{Fe}{2} ions closely follow each other.

\begin{figure}[htp!]
\centering
\includegraphics[width=0.9\linewidth]{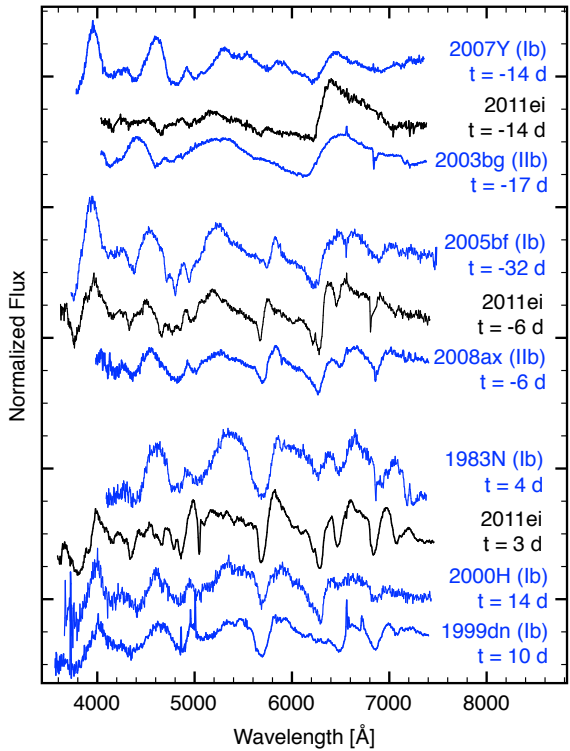}

\caption{Early spectra of SN\,2011ei compared to those of other SNe
  IIb and Ib. Spectra have been normalized according to the procedure
  outlined in \citet{Jeffery07} to aid in visual comparison.}

\label{fig:spec:earlytomax}
\end{figure}

\subsection{Comparison to SNe IIb and Ib}

SN\,2011ei shares spectral properties with many SNe IIb and
Ib. Selected examples are plotted in
Figure~\ref{fig:spec:earlytomax}. The earliest spectrum ($t=-14$~d)
shows close similarity to that of SN\,2007Y
\citep{Stritzinger09}. Both SNe exhibit a rarely-observed, extended,
high-velocity H$\alpha$ profile that dominates the spectrum. However,
SN\,2011ei does not have the broad calcium absorption seen in
SN\,2007Y around 8400 \AA, nor does it show the pronounced features
between $4000-5000$ \AA\ associated with \ion{Fe}{2} and \ion{He}{1}.
A likeness between SN\,2011ei and SN\,2003bg is also seen
\citep{Hamuy09,Mazzali09}. In this case, the H$\alpha$ profiles are
comparable, but SN 2003bg shows a blueshifted absorption associated
with higher velocities ($\sim 20,000$ \kms) and distributed
more gradually. As with SN\,2007Y, the features blueward of 5000~\AA\
are more pronounced in SN 2003bg than in SN\,2011ei.

Approximately one week before maximum light when the H$\alpha$
emission is no longer conspicuous in SN\,2011ei, its spectra resemble
those of SNe IIb from compact progenitors and SNe Ib exhibiting
absorption around $\approx 6250$~\AA. Shown in
Figure~\ref{fig:spec:earlytomax} ({\it middle section}) is the Type Ib
SN\,2005bf \citep{Folatelli06} early in its unusual late-peaking light
curve, and the Type IIb SN 2008ax \citep{Pastorello08} at $t=-6$ d.
By the time of maximum light and in the months that follow, SN\,2011ei
continues to exhibit spectral properties consistent with those of SNe
Ib. Other examples plotted in Figure~\ref{fig:spec:earlytomax} ({\it
bottom section}) are SN 1983N \citep{Harkness87}, and SNe 1999dn and
2000H (retrieved electronically via SUSPECT courtesy of the Asiago SN
Group). Of all the SNe IIb and Ib examined, the EW of the hydrogen and
helium features in the spectra of SN\,2011ei were among the most
narrow.

\begin{figure}[htp!]
\centering
\includegraphics[width=0.99\linewidth]{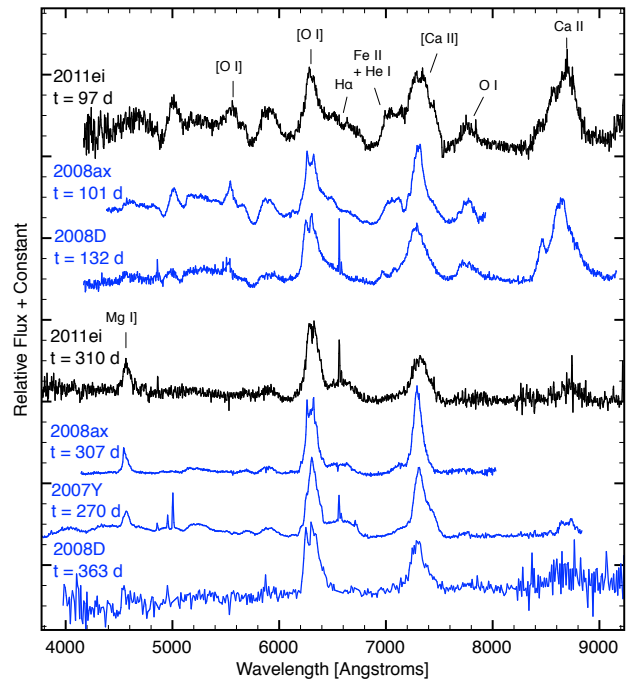}

\caption{Late-time optical spectra of SN\,2011ei compared to spectra
  of SN 2008ax \citep{Milisavljevic10}, 2007Y \citep{Stritzinger09}
  and SN 2008D (t=132 d from \citealt{Modjaz09} and t=363 d from
  \citealt{Tanaka09}). \ion{O}{1} $\lambda$8446 may contribute to
  strong emission observed around the \ion{Ca}{2} infrared triplet.}

\label{fig:latetimespec}
\end{figure}

\begin{figure}[htp!]
\centering
\includegraphics[width=0.99\linewidth]{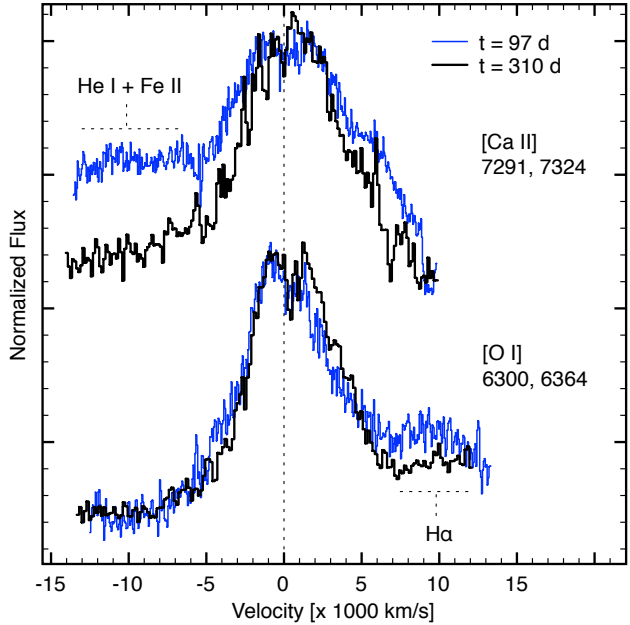}

\caption{Late-time emission line profiles of SN\,2011ei from the $t =
  97$~d and $t = 310$~d spectra. Instances where line profiles are
  contaminated by emissions from other lines are indicated.}

\label{fig:latetimeprofiles}
\end{figure}

\subsection{Nebular Phase Spectrum}
\label{sec:nebspec}

Starting with the $t = 73$ d spectrum, SN\,2011ei begins to enter the
nebular phase and emission from forbidden transitions is
apparent. Emission from the [\ion{O}{1}] \dlambda 6300, 6364,
[\ion{Ca}{2}] \dlambda 7291, 7324, and \ion{Mg}{1}] $\lambda$4571
lines originating from the inner ejecta is observed. H$\alpha$
emission attributable to high-velocity hydrogen gas interacting with
dense circumstellar material \citep{Houck96} is also present.

The overall emission properties of SN\,2011ei at nebular epochs are
quite similar to those of other SNe IIb and Ib. This is clear in
Figure~\ref{fig:latetimespec}, where the $t = 97$~d and $t=310$~d
spectra of SN\,2011ei are compared to spectra of SN\,2007Y, SN\,2008D
and SN\,2008ax.  The ratio of [\ion{Ca}{2}]/[\ion{O}{1}] emission,
which has been predicted to be a potential dianostic of progenitor
core mass \citep{Fransson89}, is relatively small in
SN\,2011ei. Unlike SN\,2008D and SN\,2008ax, SN\,2011ei does not show
conspicuous multiple peaks in its [\ion{O}{1}] \dlambda6300, 6364
emission (see, e.g.,
\citealt{Mazzali05,Modjaz08,Modjaz09,Maeda08,Taubenberger09,Milisavljevic10}).
Thus, the degree of ejecta asymmetry may be less in SN\,2011ei
compared to those events.

In Figure~\ref{fig:latetimeprofiles}, emission line profiles of the $t
= 97$~d and $t = 310$~d spectra of SN\,2011ei are enlarged and plotted
in velocity space. The fairly broad [\ion{Ca}{2}] \dlambda 7291, 7324
emission is centered roughly around zero velocity (assumed to lie at
7306 \AA). A blueward ledge that is probably a combination of
\ion{He}{1} $\lambda$7065 and [\ion{Fe}{2}] $\lambda$7155 diminishes
between the two epochs.  Close inspection reveals a non-uniform
central peak in the [\ion{O}{1}] \dlambda 6300, 6364 profile in the
form of a narrow gap centered around 6305~\AA. The reality of this
feature is certain as it appears in both epochs, and interestingly,
emission redward of the gap appears to grow stronger with time. The
observed substructure is suggestive of clumpy ejecta (see, e.g.,
\citealt{Spyromilio94}), and the increase in redshifted emission may
be associated with declining optical opacity
\citep{Taubenberger09}. Alternatively, the profile may be influenced
by the high-velocity H$\alpha$ absorption observed at early epochs
\citep{Maurer10}.

\begin{figure}[htp!]
\centering
\includegraphics[width=0.95\linewidth]{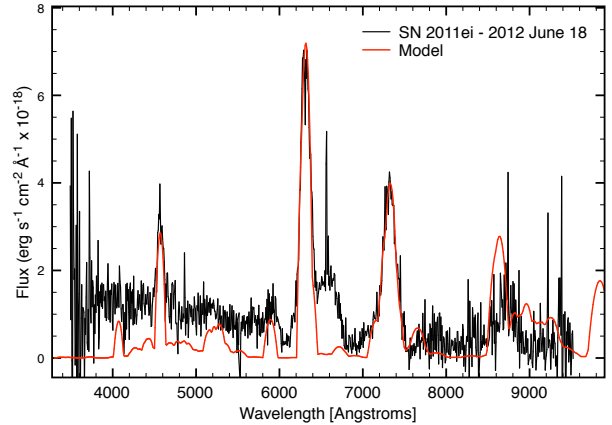}

\caption{Late-time spectrum of SN\,2011ei (black) and our nebular
  emission model (red).}

\label{fig:model}
\end{figure}

\begin{figure*}[htp!]
\begin{minipage}[b]{0.475\linewidth}
\centering
\includegraphics[width=0.98\linewidth]{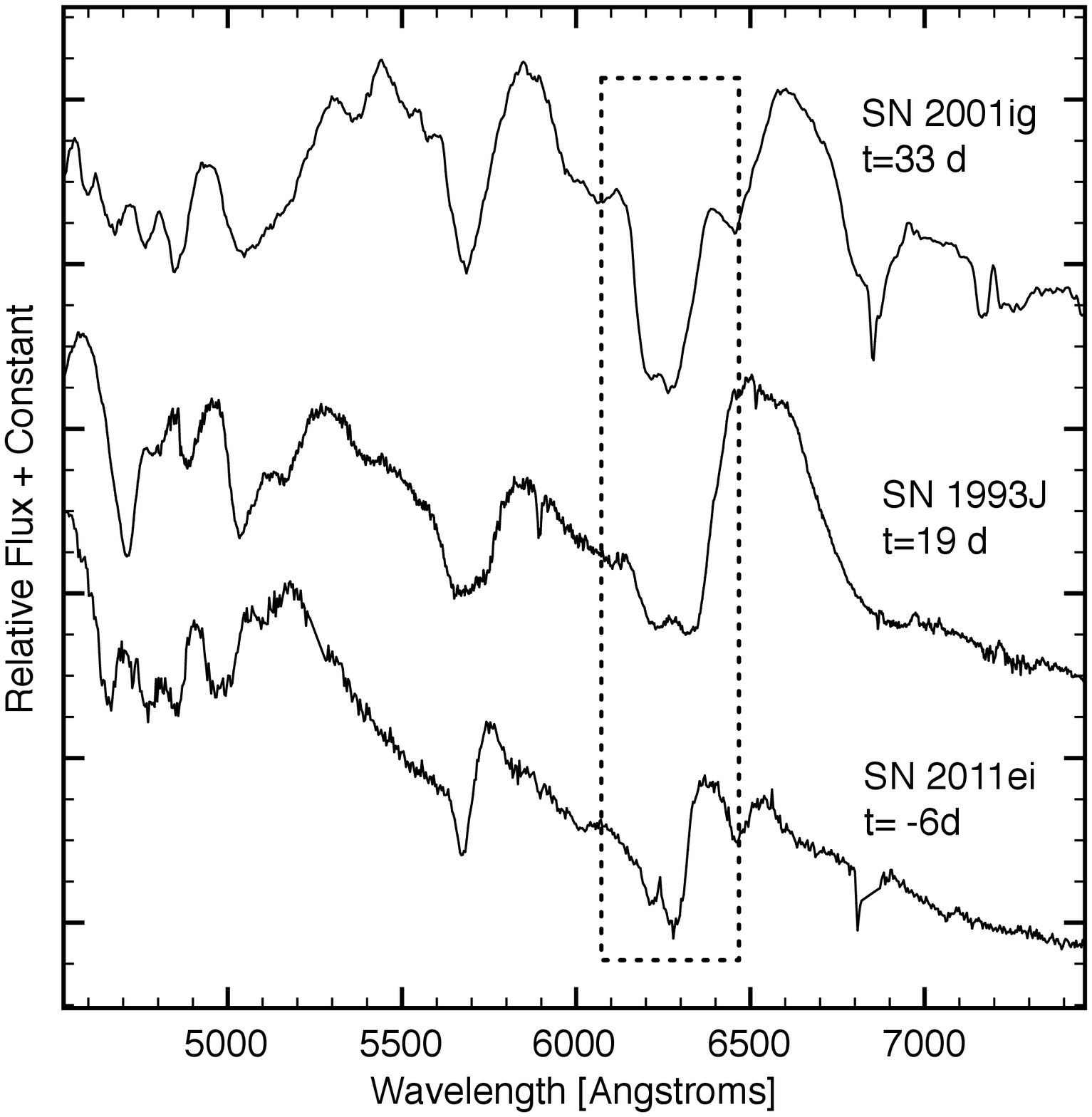}

\caption{Evidence of two-component absorption centered near $\approx
  6250$ \AA\ for SN\,2011ei, SN\,1993J \citep{Matheson00}, and
  SN\,2001ig \citep{Silverman09}. Phase is with respect to $V$-band
  maximum. The dashed box highlights the region around H$\alpha$ where
  the absorption is observed.}

\label{fig:spectra:11eiHa2compCompare}
\end{minipage}
\hspace{0.03\linewidth}
\begin{minipage}[b]{0.475\linewidth}
\centering
\includegraphics[width=0.98\linewidth]{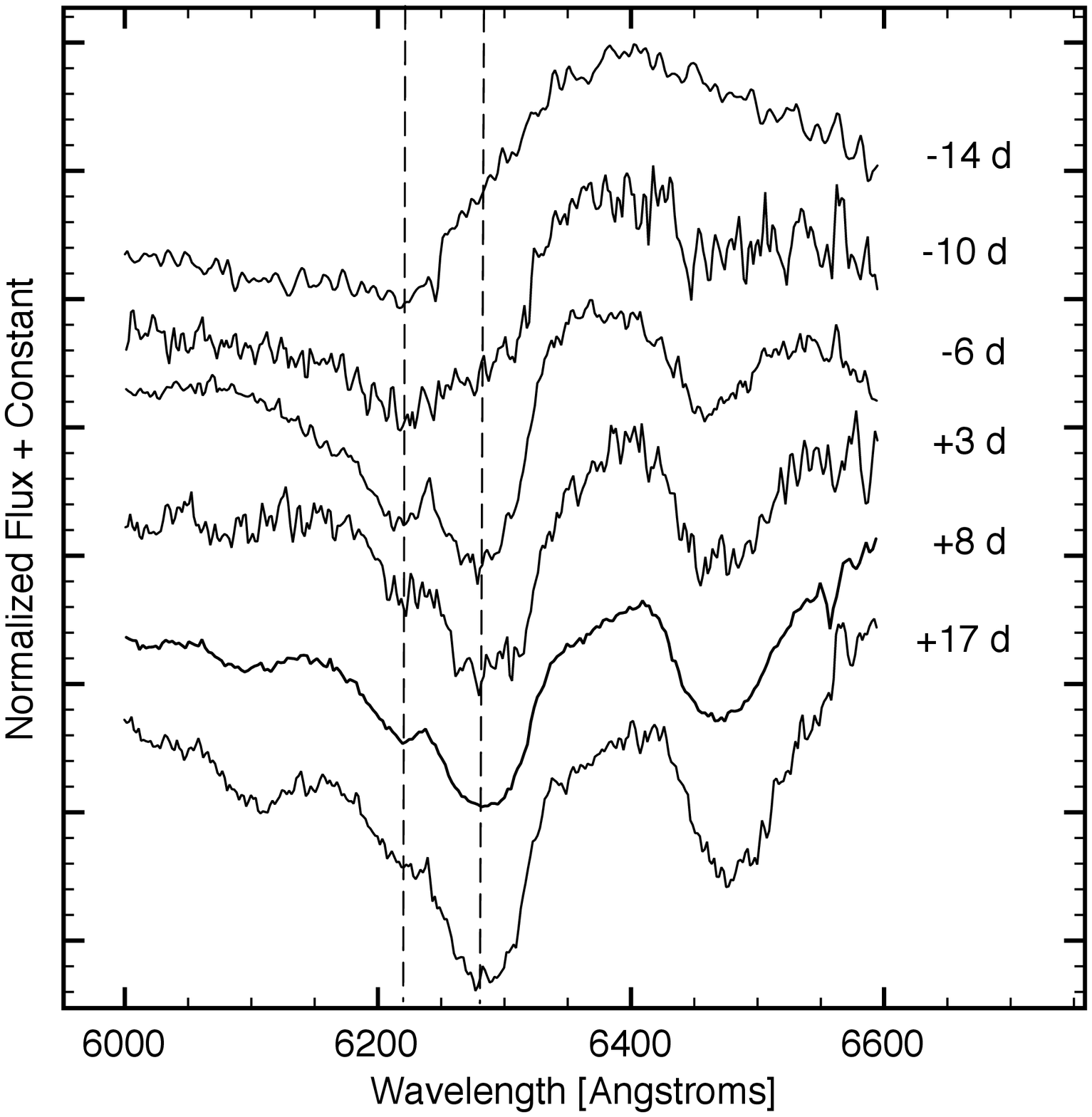}

\caption{Time series spectra of SN\,2011ei in the region of H$\alpha$
  showing evidence for two-component absorption evolution. Phase is
  with respect to $V$-band maximum. The dashed lines mark the
  individual components.}

\label{fig:Halphaevolution}
\end{minipage}
\end{figure*}

\subsection{Model of Nebular Spectrum}

We modelled the nebular spectrum of SN\,2011ei using our SN nebular
spectrum code, assuming that the late-time emission is tied to the
deposition of gamma-rays and positrons from $^{56}$Co decay. Given an
ejected mass, a characteristic boundary velocity (which corresponds to
the half-width-at-half-maximum of the emission lines), and a
composition, the code computes gamma-ray deposition, follows the
diffusions of the gamma-rays and the positrons with a Monte Carlo
scheme and computes the heating of the gas. The state of the gas is
then computed in NLTE, balancing heating and cooling via line
emission. The code has been used for a number of SNe Ibc
(e.g. \citealt{Mazzali01}) and is the latest version described in some
detail in \citet{Mazzali11}.

For SN\,2011ei, we modeled the 2012 June 18 spectrum assuming a
rest-frame epoch of $\approx 325$ days. We used a line width of 4500
\kms, a distance modulus $\mu = 32.27$, and a reddening $E(B-V)=0.24$
(see Section \ref{sec:distance}). We also assumed that the gass has a
spherical distribution, which is reasonable given that the line
profiles do not seem to indicate the need for large asymmetries (see
Section \ref{sec:nebspec}).

We found that the spectrum (excluding the H$\alpha$ line, which is
mostly the result of recombination) can be powered by a $^{56}$Ni mass
of $0.047 \pm 0.005\; M_{\odot}$. This is determined mainly by fitting
the Fe forbidden emission lines, and simultaneously the other lines.
The overall ejecta mass is quite small, $\sim 0.3\;M_{\odot}$, most of
which is oxygen ($0.21\; M_{\odot}$). Carbon, magnesium, silicon and
sulphur are also present. The Ca abundance is small ($0.002\;
M_{\odot}$) but this is sufficient to form strong emission lines.

The ejected mass we derived with the nebular model is smaller than our
other mass estimate of 1.6 $M_{\odot}$ derived from the bolometric
light curve (see Section \ref{sec:explosion}). Hydrogen can be
responsible for another $0.1\; M_{\odot}$ or so, as is typical of SNe
IIb \citep{James10,Hachinger12}, but this still means a small
mass. Possibly some of the outer O-rich ejecta are too thin to cause
the efficient deposition of gamma-rays, thus the nebular emission
model may have underestimated the mass in this way. Our estimate for
$M_{Ni}$ from nebular modeling, however, is in statistical agreement
with the estimate from the peak of the light curve, and we conclude
that the Ni mass likely lies within the range of the two estimates;
i.e., $0.03 \; M_{\odot} \lesssim M_{Ni} \lesssim 0.05 \; M_{\odot}$.

\subsection{Multi-component Absorption Due to Hydrogen}
\label{sec:dabsorption}

The absorption with two minima observed around $6200-6300$ \AA\ is not
unique to SN\,2011ei. In Figure~\ref{fig:spectra:11eiHa2compCompare},
we show spectra of SN\,2011ei, SN\,1993J and SN\,2001ig that
illustrate how these events all share broad absorption suggestive of
two components. The epoch at which this phenomenon is visible ranges
considerably from $t = 33$~d in SN\,2001ig, to $t=19$~d in SN\,1993J,
to $t=-6$~d in SN\,2011ei. In the case of SN\,2011ei, there is clear
evolution in the relative strength of the two components.  The
blueward absorption is observed first, then the second, redward
absorption appears and increases in strength (see
Figure~\ref{fig:Halphaevolution}).

The origin of the two-component feature is not clear. In SN\,1993J,
its nature has been investigated thoroughly using SYNOW spectral
fitting and NLTE modeling
\citep{Baron94,Baron95,Zhang95a,Zhang95b,Zhang96}. Suspected origins
include ejecta asymmetry, mixture with \ion{Fe}{2} or \ion{Si}{2}
lines, and a complicated two-component density structure in H-rich
material. \citet{James10} found that they were unable to model a
similar two-component feature in SN\,2000H and attributed it to
interaction with circumstellar material (CSM). They cited a
quickly-fading \ion{Na}{1} D feature observed around maximum light as
evidence for environmental interaction. More recently,
\citet{Hachinger12} attributed the feature in SNe IIb and Ib to a
blend of high-velocity H$\alpha$ and \ion{Si}{2}.

Multi-component absorption may be a common phenomenon across supernova
types and emission lines. For example, evidence for two components of
high-velocity hydrogen absorption has been seen in early spectra of
the Type Ib SN\,1999dn \citep{Benetti11}, as well as the Type IIb
SN\,2011dh during the first month following outburst (H.\ Marion,
private communication, 2012). Multi-component absorption has also be
recognized in the complex evolution of broad \ion{Si}{2} $\lambda$6355
and \ion{Ca}{2} H\&K absorption features in Type Ia supernovae
\citep{Foley11,Foley12}.

\subsection{Explosion Site Metallicity}
\label{sec:Host}

Possible metallicity differences between SN subtypes and their
relation to progenitor systems is currently an active area of
investigation (e.g.,
\citealt{Prieto08,Arcavi10,Anderson10,Modjaz11,Kelly11,Sanders12}).
We were therefore motivated to estimate the metallicity of the host
environment of SN\,2011ei. 

In our 2012 June 18 spectrum of SN~2011ei we detect narrow nebular
emission lines of H$\alpha$, H$\beta$, [\ion{N}{2}] $\lambda6583$, and
[\ion{O}{2}] $\lambda$3727.  From the N2 diagnostic of \citet{PP04},
we estimate an oxygen abundance of $\log({\rm O/H})+12=8.8\pm0.1$.
Using the N202 diagnostic of \cite{Kewley02}, we measure $\log({\rm
O/H})+12=9.0\pm0.1$.  These two values are consistent given the well
known offset between the diagnostics \citet{KE08}.

Adopting a solar metallicity of $\log({\rm O/H})_\odot+12=8.69$
\citep{Asplund05} on the PP04 scale, our measurement suggests that the
environment of SN~2011ei has an approximately solar metallicity. This
puts SN\,2011ei on the high end of the observed distribution of SN IIb
explosion site metallicities (see, e.g.,
\citealt{Kelly11,Modjaz11,Sanders12}).

\begin{figure*}[htp!]
\begin{minipage}[b]{0.475\linewidth}
\centering
\includegraphics[width=\linewidth]{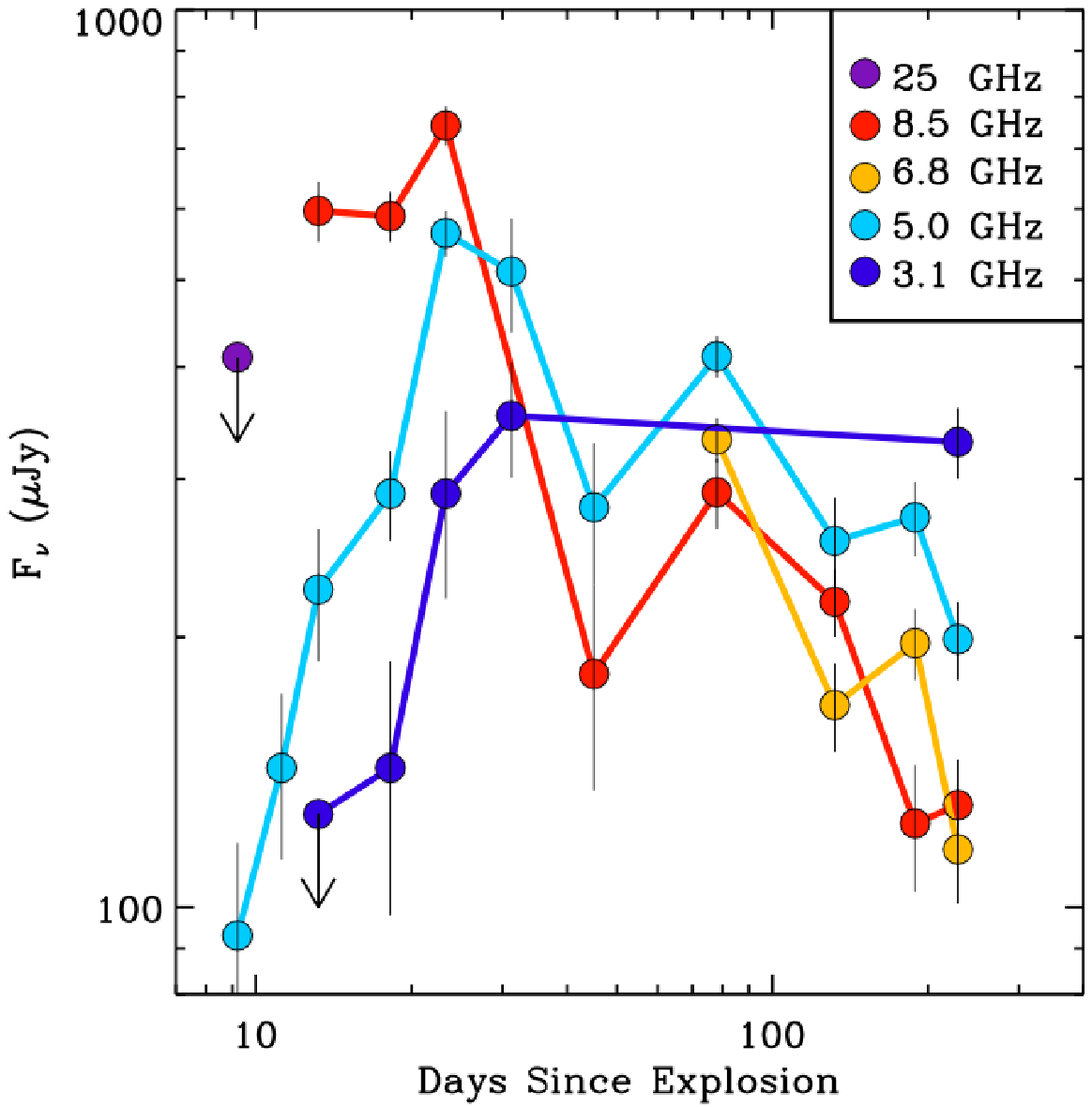}

\caption{VLA radio light curves of SN\,2011ei are shown at frequencies
  spanning 3.1 to 25 GHz.  At the best sampled frequency, 5 GHz, we
  find evidence for a secondary maximum at $\Delta t_{\rm exp}\approx
  80$ days that is also seen in the 8.4 GHz light curve.}

\label{fig:lt_curves}
\end{minipage}
\hspace{0.04\linewidth}
\begin{minipage}[b]{0.475\linewidth}
\centering
\includegraphics[width=\linewidth]{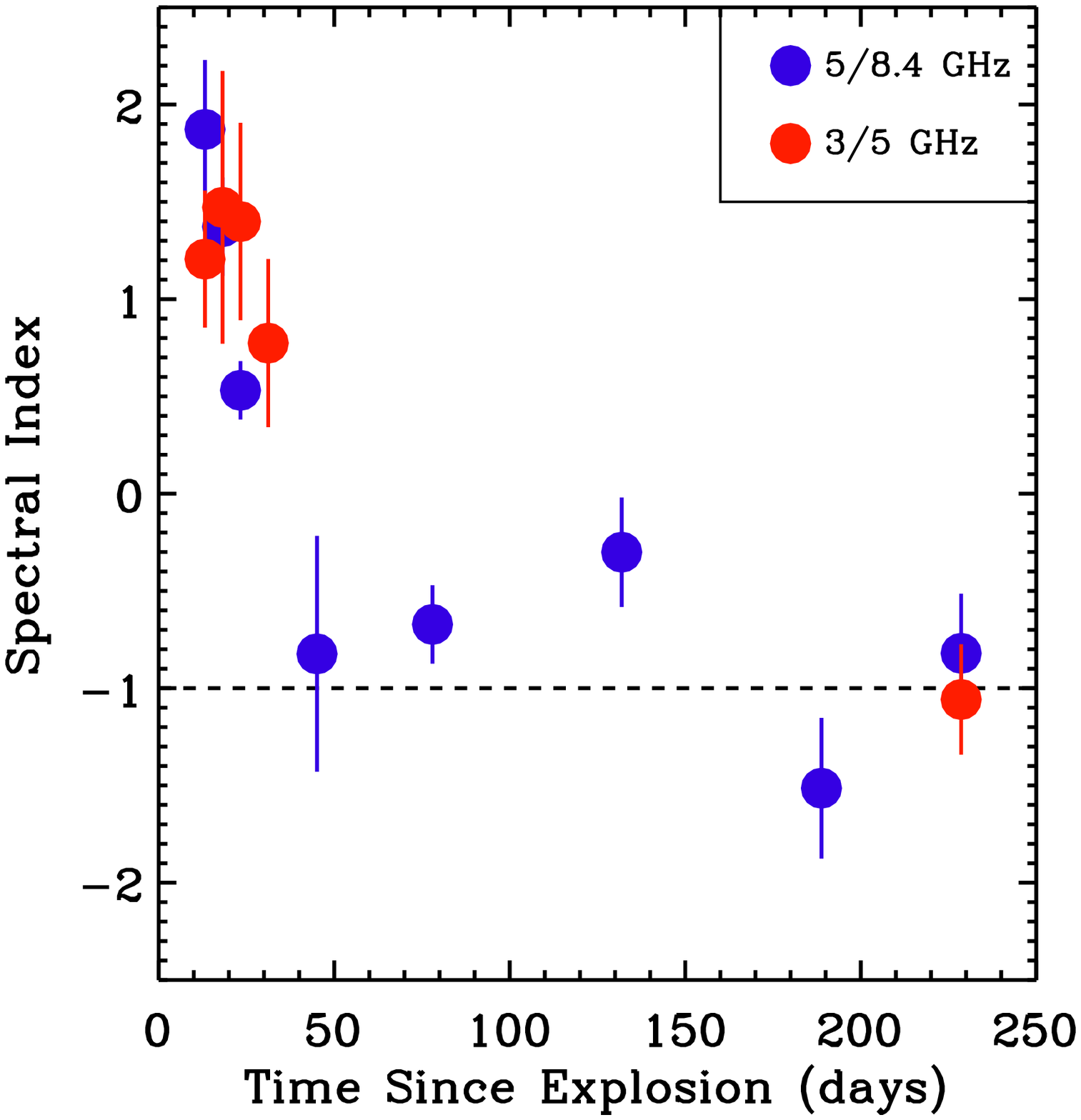}

\caption{The evolution of the radio spectral indices spanning 3 to 8.4
GHz are shown.  The early radio emission is optically thick,
$F_{\nu}\propto \nu^{2}$, but gradually evolves to an optically thin
spectrum, $F_{\nu}\propto \nu^{-1}$.}

\label{fig:spectral_indices}
\end{minipage}
\end{figure*}


\section{Radio and X-ray Diagnostics}
\label{sec:diagnostics}


\subsection{Radio Light Curves, Forward Shock, and
                    Magnetic Field}

While optical data probe the thermal emission from the slow-moving
bulk ejecta, non-thermal synchrotron emission in the radio is
produced as the forward shock races ahead and dynamically interacts
with the CSM. Radio observations can thus provide unique information
about the progenitor star's mass-loss history several years to
centuries immediately prior to outburst.

We show the observed radio light curves of SN\,2011ei in
Figure~\ref{fig:lt_curves}. The radio light curves show flux density
variations of a factor of $\sim 3$ with a 5 GHz peak spectral
luminosity, $L_{\nu}\approx 5.48\times 10^{26}~\rm
erg~s^{-1}~Hz^{-1}$, approximately 23 days after the explosion.  The
peak radio emission is fairly weak compared to the majority of SNe IIb
and Ibc, but comparable to, e.g., SN\,2011dh, SN\,2008ax, and
SN\,1996cb \citep{Soderberg11,Krauss12,Chevalier10}.  Approximately 40
days after the explosion, the radio emission decays steeply at all
frequencies, then rises again to a second maximum $\sim 2$ months
later.

As the forward shock expands into the surrounding environment,
electrons are accelerated to relativistic velocities with a
distribution, $N(\gamma_e)\propto \gamma_e^{-p}$, where $\gamma_e$ is
the Lorentz factor of the particles.  Amplified magnetic fields cause
the shocked electrons to gyrate and emit synchrotron emission
detectable in the cm-band on a timescale of days after the explosion
\citep{Chevalier82}.  A low frequency spectral turnover has been
observed for radio SNe Ibc and IIb attributed to synchrotron
self-absorption defining a spectral peak, $\nu_a$, and the spectrum is
characterized by $F_{\nu}\propto \nu^{5/2}$ below $\nu_a$ and
$F_{\nu}\propto \nu^{-(p-1)/2}$ above $\nu_a$.  At the epoch $\nu_a$
crosses each observing band(s), the radio light curve reaches maximum
intensity.

In Figure~\ref{fig:spectral_indices}, we show the radio spectral
indices in adjacent bands spanning 3 to 8 GHz for SN\,2011ei.  The
early radio emission is optically thick with $F_{\nu}\propto \nu^{2}$ while
the later emission is optically thin with approximately
$F_{\nu}\approx \nu^{-1}$, implying $p\approx 3$. As shown in
Figure~\ref{fig:lt_curves}, the 5 GHz light curve peaks on 2011 August
17 ($\Delta t_{\rm exp} \approx 23$ days) with a flux density of
$F_{\nu,p}\approx 560~\mu$Jy.  Adopting $p=3$, we estimate the
time-averaged velocity, $\overline{v}\equiv R/\Delta t$, of the
forward shock to be,
\begin{eqnarray}
\overline{v}\approx 0.14c &\times&(\epsilon_e/\epsilon_B)^{-1/19}
(F_{\nu,p}/~\rm mJy)^{9/19}\nonumber \\
&\times&(d/10~{\rm Mpc})^{18/19} (\nu_p/5~{\rm GHz})^{-1}\nonumber\\
&\times&(t_p/10~{\rm days})^{-1},
\end{eqnarray}
or $\overline{v}\approx 0.13\,c$ (see \citealt{CF06}).

Here we have assumed that half of the volume enclosed by the forward
shock is producing synchrotron emission\footnote{Very long baseline
interferometry radio images of SN\,1993J suggested a spherical shell
structure with a shell thickness 20-33\% of the outer radius
\citep{Bietenholz+2011,Marti-Vidal+2011a}, in which case 49-70\% of
the volume enclosed by the forward shock is filled with radio
emission. However, possible small scale clumpiness within this
geometry could further lower the effective radio-emitting volume.}, and
$\epsilon_e$ and $\epsilon_B$ represent the efficiency of the shock
wave in accelerating electrons and amplifying magnetic fields,
respectively.  We have further assumed that the shock is in
equipartition (defined as $\epsilon_e=\epsilon_B$).  The inferred
shock wave velocity for SN\,2011ei is in line with the velocities
derived for SNe Ibc and compact progenitor SNe IIb
\citep{Chevalier10,Soderberg11}, and higher than those inferred for
extended progenitor SNe IIb and SNe of Type IIP \citep{Chevalier98}.

Following \citet{CF06}, we find the strength of the amplified magnetic
field ($B$) for the radio emitting material to be
\begin{eqnarray}
B\approx 0.68 &\times&
(\epsilon_e/\epsilon_B)^{-4/19} (F_{\nu,p}/~\rm
mJy)^{-2/19}\nonumber\\
&\times&(d/10~{\rm Mpc})^{-4/19} (\nu_p/5~{\rm GHz})\;{\rm G},
\end{eqnarray}
or $B\approx 0.6$ G for equipartition.

\begin{figure}[htp!]
\centering

\includegraphics[width=\linewidth]{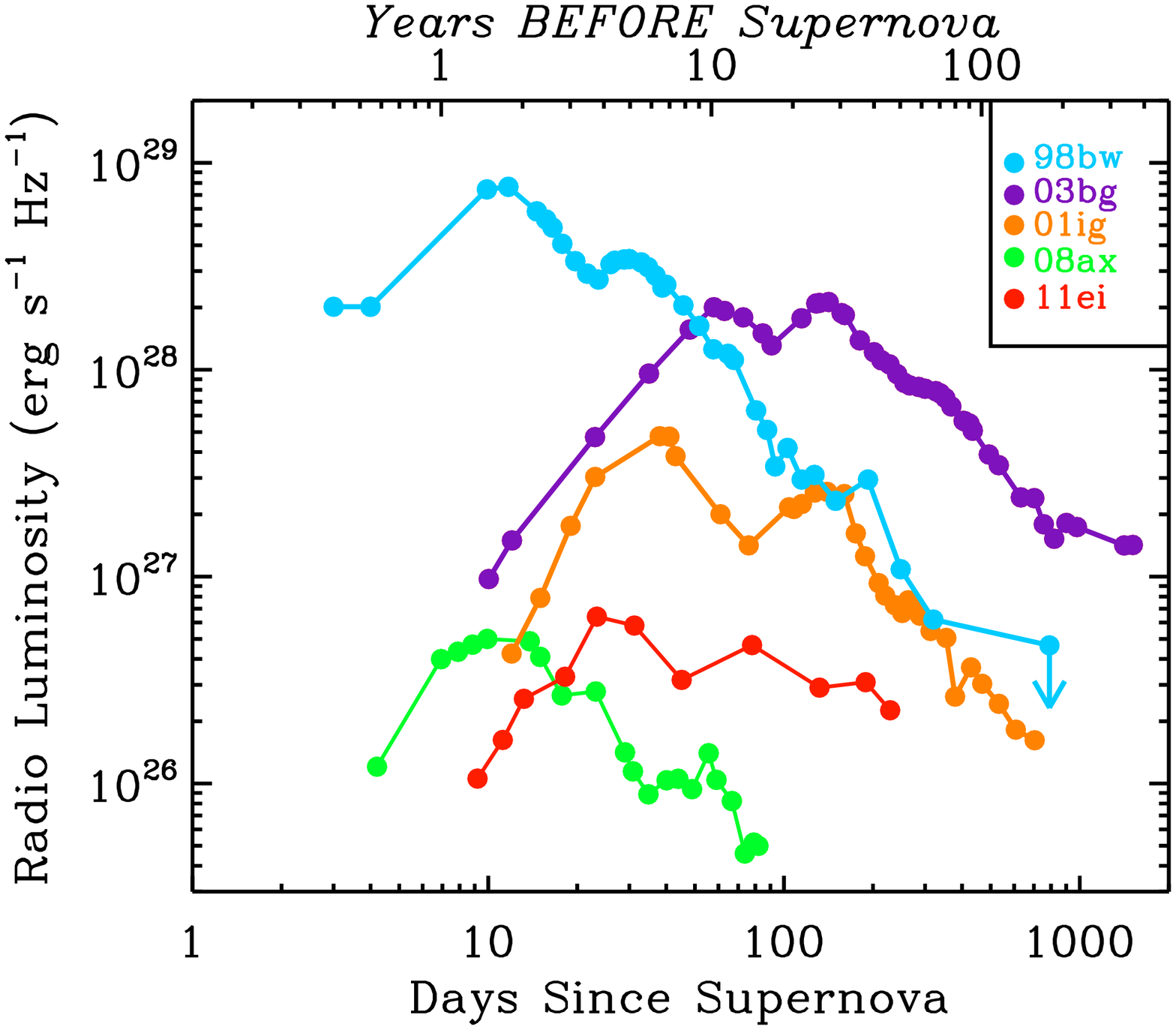}

\caption{The 5 GHz light curve for SN\,2011ei is compared to those of
  other radio supernovae including the Type IIb SNe 2001ig
  \citep{Ryder04}, SN\,2003bg \citep{Soderberg06}, SN\,2008ax
  \citep{Roming09}, and the GRB-associated Type Ic SN\,1998bw
  \citep{Kulkarni98}.  Bottom $x$-axis represents days since supernova
  outburst, and the top $x$-axis represents inferred times of CSM
  density modulations in years previous to outburst (assuming
  $v_{w}=1000$ \kms). Each of these SNe shows variable radio emission
  that deviates from the expected decay as roughly $F_{\nu}\propto
  t^{-1}$.  In particular, all of these events show second maxima
  within a few months of the explosion.}

\label{fig:lum_limits_2011ei}

\end{figure}

\subsection{Mass-loss Rate and Evidence for an Inhomogeneous Wind}
\label{sec:radio}

Finally, we consider the mass-loss rate of the progenitor system.  As
shown in Figure~\ref{fig:lt_curves}, the radio light curves for
SN\,2011ei do not decline as a power-law following maximum light, but
instead show evidence for a secondary maximum at $\approx 80$ days
post-outburst.  As noted in \citet{Chevalier10}, secondary maxima are
common among SNe IIb.  Assuming the dominant absorption process is
synchrotron self-absorption as suggested by
Figure~\ref{fig:spectral_indices}, the time averaged mass-loss rate
$\dot M$ from the progenitor star can be estimated as
\begin{eqnarray}
\dot M &\approx& 0.38\;M_{\odot}~{\rm yr^{-1}}\nonumber\\
  &\times&(\epsilon_B/0.1)^{-1} (e_e/e_B)^{-8/19}(F_{\nu,p}/~\rm
  mJy)^{-4/19}\nonumber\\
  &\times&(d/10~\rm Mpc)^{-8/19}(\nu_p/5)^{2} (t_p/10~\rm days)^{2},
\end{eqnarray}
for a wind velocity of $v_w=1000~\rm km~s^{-1}$ typical for a
WR star \citep{CF06}.  

For reasonable values of $\epsilon_e=e_B=0.1$ we find $\dot M \approx
1.4 \times 10^{-5}\;M_{\odot}\,\rm yr^{-1}$.  We note that this is a
lower bound on the inferred mass-loss rate from the progenitor
system. Lower values of the partition fractions and deviations from
equipartition (see, e.g., SN\,1993J; \citealt{Fransson98}) only serve
to increase $\dot M$.  The mass-loss rate for SN\,2011ei is similar to
within an order of magnitude to that inferred for other SNe IIb and
Ibc.

As seen in Figure~\ref{fig:lum_limits_2011ei}, radio light curve
variations observed similar to those observed in SN\,2011ei have been
detected for other SNe IIb including SN\,2001ig \citep{Ryder04} and
SN\,2008ax \citep{Roming09}, the broad-lined event SN\,2003bg
\citep{Soderberg06}, as well as the GRB-associated SN\,1998bw
\citep{Kulkarni98}. These fluctuations are not dissimilar in both
timescale and amplitude, and have been reasonably explained in terms
of density modulations in the pre-explosion environment shaped by the
progenitor system \citep{Ryder04,Soderberg06,Roming09}.

We adopt a model in which the modulations in the radio light curve are
similarly due to circumstellar density variations.  As shown in
\citet{Wellons12}, the optically-thin synchrotron emission scales with
the number density of shocked electrons as $F_{\nu}\propto n_e^2$ for
$p=3$.  Thus, assuming a constant compression factor by the forward
shock, we find that the circumstellar density modulations scale
similarly.  Given the factor of $\sim 3$ in flux density modulations
observed for SN\,2011ei, we infer a factor of $\sim 2$ jump in the
circumstellar density at a radius $r\sim 2\times 10^{16}$ cm.
Attributing this effect to a variation in the progenitor wind of the
progenitor star implies an ejection timescale of $\sim
7\;(v_w/10^3~\rm km~s^{-1})$ yr prior to outburst.

It is intriguing to note that radio modulations have also been
observed in relativistic, engine-driven SNe including GRB-SN\,1998bw
\citep{Kulkarni98,Li99} and SN\,2009bb \citep{Bietenholz10}.  In these
cases, the flux density modulations have been attributed to energy
injection from the central engine.  However, given the resemblance to
the observed compact progenitor Type IIb radio modulations, we
speculate that CSM density fluctuations on radial scales of $\lesssim
10^{17}\;\rm cm$ may be common among stripped-envelope SN explosions.

\subsection{Limits on Inverse Compton X-ray Emission}

For hydrogen-stripped SNe exploding in low density environments, the
dominant X-ray emission mechanism during the first weeks to a month
after the explosion is Inverse Compton (IC; \citealt{BjFr04,CF06}). In
this framework, X-ray photons originate from the up-scattering of
optical photons from the SN photosphere by a population of electrons
accelerated to relativistic speeds by the SN shock. X-ray IC depends
on the density structure of the SN ejecta, the structure of the CSM,
and the details of the electron distribution responsible for the
up-scattering, but does not require any assumption on magnetic field
related parameters and it is not affected by possible uncertainties on
the SN distance. Thus, limits on X-ray emission can provide
information on the pre-SN mass-loss history of the progenitor
independent of our radio analysis.

Using the reconstructed bolometric luminosity and the derived
explosion parameters ($M_{ej} \sim 1.6\;{M_{\sun}}$ and $E_{k} \sim
2.5\times 10^{51}\;\rm{erg}$), and adopting the formalism discussed in
\cite{Margutti12}, we estimated the progenitor mass-loss rate via
X-ray IC. We assumed (i) the fraction of energy going into
relativistic electrons was $\epsilon_e=0.1$ as indicated by well
studied SN shocks \citep{CF06}, (ii) a power-law electron distribution
$N(\gamma_e)=n_0 \gamma_e^{-p}$ with $p=3$ as indicated by radio
observations of SNe Ibc \citep{CF06}, (iii) a wind-like CSM that
follows $\rho_{\rm{CSM}}\propto R^{-2}$ as expected from a star which
has been losing material at constant rate $\dot M$, and (iv) that the
outer density structure of the ejecta scales as $\rho_{\rm{SN}}\propto
R^{-n}$ with $n\sim10$ (see e.g., \citealt{Matzner99,CF06}). Given
these assumptions, the \emph{Chandra} upper limit on X-ray emission of
$7.6\times 10^{-4}\,\rm{c\,s^{-1}}$ in the 0.5-8 keV band implies
$\dot M <2\times10^{-4}\; M_{\sun}\; \rm yr^{-1}$, for wind velocity
$v_w = 1000\;\rm km\,s^{-1}$.  This calculation is consistent with the
properties derived from our radio analysis (Section \ref{sec:radio}).


\section{Discussion}
\label{sec:Discussion} 


\subsection{The Progenitor of SN 2011ei}
\label{sec:progenitor}

The early onset of helium-rich features in the optical spectra of
SN\,2011ei and the evolution of its bolometric light curve are
consistent with model explosions of He core stars with pre-supernova
masses of $\sim 3-4\;M_{\odot}$ \citep{Shigeyama90,Woosley95}. The
initial presence and subsequent rapid disappearance of hydrogen
suggests the progenitor star still retained a thin hydrogen envelope
of mass $< 0.1\;M_{\odot}$ immediately prior to outburst. 

Our inferred Ni mass of $\approx 0.04 \;M_{\odot}$ for SN\,2011ei is
on the low end of Ni masses derived for other stripped-envelope events
($\sim 0.05-0.15\;M_{\odot}$; \citealt{Taubenberger11}). A low Ni mass
yield is consistent with the explosions of lower-mass He star
progenitors that have small iron cores ($\la 0.15\;M_{\odot}$) and
eject considerably less material than do their larger counterparts
\citep{Shigeyama90,Hachisu91}.  Lower-mass He stars are also predicted
to undergo more extensive mixing than higher-mass stars do
\citep{Hachisu91}. The helium and iron expansion velocities determined
from our synthetic spectral fits to SN\,2011ei are quite similar (see
Figure~\ref{fig:velocities}), and thus are in agreement with the
emitting ejecta being well-mixed.

Several properties of WR stars make them a plausible candidate
progenitor for SN\,2011ei: (i) WR stars are believed to be likely
progenitors of at least some SNe IIb and Ibc \citep{Heger03}, (ii)
some subtypes of WR stars (e.g., the WN class) have an observable
amount of hydrogen at their surface \citep{Hamann91}, (iii) the
radio-derived mass-loss rate of SN\,2011ei is consistent with the
observed rates for Galactic WR stars ($10^{-4} \la \dot M \la
10^{-6}\;M_{\odot}$ yr$^{-1}$; \citealt{Hamann06}), and (iv) the
progenitor radius inferred from the early photometry ($R_* \lesssim 1
\times 10^{11}$ cm) is consistent with those of WR stars
\citep{Crowther07}.  The putative WR star could have evolved
originally from either a single star with a high main sequence mass of
$\sim 25\;M_{\odot}$, or a lower-mass star with main sequence mass
$\sim 10 - 15\;M_{\odot}$ in an interacting binary system. 

The physical origin of the radio light curve variations and the
inferred inhomogeneous wind is not clear. Interestingly, not all radio
SNe show this behavior. For instance, several well-studied examples
exhibit smooth light curve evolutions, including the Type IIb SNe
1993J and 2011dh \citep{vanDyk94,Bartel+2002,Krauss12}, and the Type
Ic SN\,2003L \citep{Soderberg05}, among others. Thus, some aspect of
the progenitors of stripped-envelope events like SN\,2011ei (see
Figure~\ref{fig:lum_limits_2011ei}) produces variability in the winds
of the progenitor systems on the timescale of $\sim 10$ yr.

Luminous blue variables (LBVs) have been suggested as possible
progenitor stars given that they share many characteristics of WR
stars and can undergo occasional giant eruptive events that play a
major role in removing the H-rich envelope on timescales that are
consistent with SN\,2011ei's radio light curve variability (see, e.g.,
\citealt{Kotak06}). However, SNe IIb and Ib progenitors appear to be
less massive and more compact than LBVs
\citep{Soderberg06}. Ultimately, although LBVs may not represent the
evolutionary state immediately prior to outburst, LBV-like eruptions
may provide the dominant method of stripping the H-rich envelope at
low metallicity for WR stars \citep{Crowther07}.

\subsection{Classification Confusion: IIb vs.\ Ib}

At the earliest epochs observed, SN\,2011ei exhibited emission
properties consistent with a Type II classification.  However, by the
time of maximum light, SN\,2011ei closely resembled many SNe Ib (see
Figure \ref{fig:spec:earlytomax}). If the first optical spectra of
SN\,2011ei had only been obtained near maximum light, i.e., about one
week after we in fact obtained the first spectrum, its early H$\alpha$
emission would have been missed and the SN likely classified as a Type
Ib event.

Two important lessons come from this rapid spectral evolution. One is
that absorption around 6250 \AA\ in SNe Ib can be interpreted as a
clear signature of hydrogen.  High-velocity hydrogen has long been
suspected in SNe Ib spectra, but direct identification has been
largely circumstantial because it has most often only been observed as
absorption in the region of H$\alpha$. However, there are now at least
two other well-observed cases where the transition of an H$\alpha$
profile showing broad emission has been monitored through to
conspicuous absorption around 6250~\AA: SN\,2007Y
\citep{Stritzinger09}, and SN\,2008ax \citep{Chornock11}.

The second lesson is that the rapid disappearance of H$\alpha$
emission in the early spectra of SN\,2011ei underscores a
classification ambiguity. Namely, {\it in some cases, whether an event
is classified as Type Ib or IIb depends on how early the first
spectrum is obtained}. Consequently, criteria that take into account
the possibility of temporal selection effects should be developed to
distinguish between SNe IIb and Ib. An unambiguous classification
scheme is crucial if precise rates of SNe IIb vs.\ Ib events are to be
estimated.

\begin{figure}
\centering
\includegraphics[width=\linewidth]{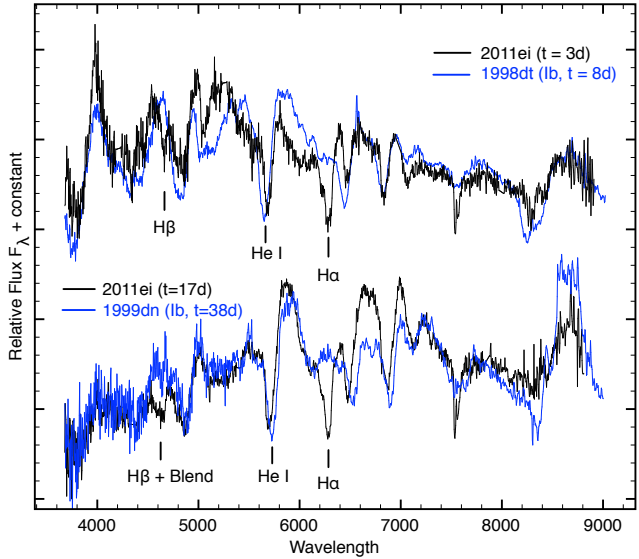}\\

\caption{Spectra of SN\,2011ei and the Type Ib SNe 1998dt and 1999dn
  \citep{Matheson01}. The relative strength of \ion{He}{1}
  $\lambda$5876 absorption (observed near 5700 \AA) to that of
  H$\alpha$ absorption (observed near 6250 \AA) could be used as a
  diagnostic to distinguish between Type IIb and Ib events.}

\label{fig:spectra:11ei_hydrogen_comparisons}
\end{figure}

One possible spectral diagnostic for distinguishing between SNe IIb
and Ib may be the relative strength of the two absorption features
around 6250 and 5700 \AA. In
Figure~\ref{fig:spectra:11ei_hydrogen_comparisons}, spectra of
SN\,2011ei and the Type Ib SN\,1998dt \citep{Matheson01} obtained
around the time of maximum light are plotted. Both share many similar
spectral features outside of the conspicuous H$\alpha$ absorption only
observed in SN\,2011ei. Also plotted in
Figure~\ref{fig:spectra:11ei_hydrogen_comparisons} are later-epoch
spectra of SN\,2011ei and the prototypical Type Ib SN\,1999dn
\citep{Matheson01}. Again, all spectral features are largely identical
outside of absorptions around H$\alpha$ and H$\beta$ only seen in
SN\,2011ei. This pattern is observed in other SNe IIb as well; e.g.,
SN\,1993J and SN\,2001ig (see
Figure~\ref{fig:spectra:11eiHa2compCompare}). Thus, a Type IIb event
may be distinguished from a Type Ib by an H$\alpha$ absorption comparable
to \ion{He}{1} $\lambda$5876 around the time of maximum light. This
could be defined explicitly as EW(H$\alpha$)/EW(\ion{He}{1}
$\lambda$5876) $\ga 0.5$ within a few weeks of maximum.


\section{Conclusions}
\label{sec:Conclusions}


We have presented X-ray, UV/optical, and radio observations of the
He-rich, stripped-enveloped, core-collapse SN\,2011ei beginning within
$\sim 1$ day of explosion. The key findings of our analyses can be
summarized as follows:

\begin{my_enumerate}

\item {\it SN\,2011ei was a relatively faint supernova with properties
    consistent with the outburst of a He-core star that suffered
    modest episodic mass-loss.}  SN\,2011ei was one of least luminous
    stripped-envelope SNe observed to date. Our estimate of its
    $^{56}$Ni mass (between $0.03$ and $0.05 \; M_{\odot}$) places it
    on the extreme low end of $^{56}$Ni masses estimated for other SNe
    Type IIb and Ibc. Radio light curve fluctuations in SN\,2011ei
    suggest punctuated mass-loss episodes prior to outburst. These
    inhomogeneities are similar to a subset of other radio SNe and
    complicate the standard picture for smooth radiation-driven
    stellar winds. We suggest that a $3-4\;M_{\odot}$ WR star of
    radius $R_* \la 1 \times 10^{11}$~cm that retained a thin hydrogen
    envelope immediately prior to outburst as a plausible candidate
    progenitor for SN\,2011ei.

\item {\it High-velocity hydrogen is not rare in SNe Ib.}
  Identification of hydrogen in Type Ib spectra has been largely
  circumstantial because it has most often only been observed as
  absorption around H$\alpha$. In SN\,2011ei, however, the transition
  of hydrogen emission through to absorption was closely monitored to
  identify its origin unambiguously. This evolution supports notions
  of a spectroscopic sequence bridging SNe IIb, SNe Ib that have deep
  H$\alpha$ absorptions, and typical SNe Ib via a common progenitor
  scenario potentially rooted in differences in the hydrogen envelope
  mass. Whether this has origins in binary interaction and/or strong
  stellar winds is not clear, but circumstantial evidence suggests
  stellar duplicity plays a role in at least some cases.

\item {\it Time-dependent classifications of SNe IIb and Ib bias
    estimates of their explosion rates.} Like SN\,2007Y
    \citep{Stritzinger09} and SN\,2008ax \citep{Chornock11},
    SN\,2011ei was caught early enough to observe a rapid evolution
    from Type II to Type Ib features in its pre-maximum light spectra.
    While SNe IIb have traditionally been understood to undergo this
    transformation on the timescale of months, examples such as
    SN\,2011ei establish that the metamorphosis can occur on the
    timescale of a few days. Consequently, how close an observation is
    made relative to the time of explosion is a significant factor in
    a Type IIb vs.\ Ib classification, and this has implications in
    determining precise rates.

\end{my_enumerate}

Study of SN\,2011ei shows that defining SNe Ib solely on the absence
of hydrogen can lead to a classification confusion. There is no clear
separation of SNe IIb vs.\ Ib for at least a subset of these
events. Instead, it is a gradual transition depending on remaining
hydrogen mass. Along similar lines, the current definition of SNe Ic
as lacking both hydrogen and helium lines may require revision. It is
common practice to model SNe Ic in terms of core-collapse in bare
carbon-oxygen cores \citep{Iwamoto94,Foley03,Mazzali04}. However,
there have been many reports of hydrogen and helium in SNe Ic
\citep{Filippenko88,Filippenko92,Filippenko90,Jeffery91,Branch02,Branch06},
and many objects exhibit properties that bridge SNe Ib and Ic
subtypes; e.g., SN\,2008D \citep{Soderberg08}, and SN\,2009jf
\citep{Valenti11}.

Thus, similar temporal selection effects may be at the root of
uncertain line identifications in SNe Ib and Ic. We recommend that
multi-wavelength data suites like the one presented here could help
resolve these ambiguities. Investigations of new SNe Ibc should
incorporate high-cadence ($\sim 2$ day) spectroscopic and photometric
monitoring in the UV/optical commencing within days of explosion, and
be coordinated with follow-up X-ray and radio observations. From these
panchromatic data, unique information about the progenitor star's
poorly understood evolutionary state and associated mass-loss in the
years immediately prior to SN outburst can be extracted.  Accumulation
of even just a handful of additional objects studied in this way could
establish fresh new insights into the nature of stripped-envelope SN
explosions.

\acknowledgements

We thank an anonymous referee for suggestions that improved the
manuscript. R.\ Chornock provided helpful comments. Some of the
observations reported in this paper were obtained with the Southern
African Large Telescope (SALT), as well as the 6.5 meter Magellan
Telescopes located at Las Campanas Observatory, Chile. Additional
observations were obtained at the Southern Astrophysical Research
(SOAR) telescope, which is a joint project of the Minist\'{e}rio da
Ci\^{e}ncia, Tecnologia, e Inova\c{c}\~{a}o (MCTI) da Rep\'{u}blica
Federativa do Brasil, the U.S. National Optical Astronomy Observatory
(NOAO), the University of North Carolina at Chapel Hill (UNC), and
Michigan State University (MSU). Support for this work was provided by
the National Aeronautics and Space Administration through Chandra
Award Number 12500613 issued by the Chandra X-ray Observatory Center,
which is operated by the Smithsonian Astrophysical Observatory for and
on behalf of the National Aeronautics Space Administration under
contract NAS8-03060. The National Radio Astronomy Observatory is a
facility of the National Science Foundation operated under cooperative
agreement by Associated Universities, Inc.  L.~C.\ is a Jansky Fellow
of the National Radio Astronomy Observatory. G.~P.\ acknowledges
support from ``proyecto regular UNAB'' DI-28-11/R. All SAAO and SALT
co-authors acknowledge the support from the National Research
Foundation (NRF) of South Africa. A.~B.\ was supported by a Marie
Curie Outgoing International Fellowship (FP7) of the European Union
(project number 275596). G.~P., F.~B., and J.~A.\ acknowledge support
provided by the Millennium Center for Supernova Science through grant
P10-064-F (funded by ``Programa Bicentenario de Ciencia y Tecnologia
de CONICYT'' and ``Programa Iniciativa Cientifica Milenio de
MIDEPLAN''). F.~B. acknowledges support from FONDECYT through
Postdoctoral grant 3120227. J.~A.\ acknowledges support by CONICYT
through FONDECYT grant 3110142, and by the Millennium Center for
Supernova Science (P10-064-F), with input from the ``Fondo de Innvacin
para La Competitividad, de Ministerio de Economa, Fomento Y Turismo de
Chile.''  Partially based on observations (program ID 184.D-1140)
collected at the European Organisation for Astronomical Research in
the Southern Hemisphere, Chile. S.~B.\ is partially supported by the
PRIN-INAF 2009 with the project ``Supernovae Variety and
Nucleosynthesis Yields.''  ESO VLT observations were taken within the
GO program 089.D-0032 (PI: P.\ Mazzali) with UT1/FORS2. P.~M.\ and
E.~P.\ acknowledge financial support from INAF PRIN 2011 and ASI-INAF
grants I/088/06/0 and I/009/10/0. D.~M, E.~P., and P.~M.\ thank the
Institute of Nuclear Theory at University of Washington where part of
this work was accomplished. This paper made extensive use of the
SUSPECT database (\texttt{http://www.nhn.ou.edu/$\sim$suspect/}). IRAF is
distributed by the National Optical Astronomy Observatory, which is
operated by the Association of Universities for Research in Astronomy
(AURA) under cooperative agreement with the National Science
Foundation. PYRAF is a product of the Space Telescope Science
Institute, which is operated by AURA for NASA. Dust maps were accessed
via the NASA/IPAC Infrared Science archive at
http://irsa.ipac.caltech.edu/applications/DUST/.

{\it Facilities:} \facility{SALT (RSS)},
\facility{Magellan:Baade (IMACS)},
\facility{Magellan:Clay (LDSS3)}
\facility{SWIFT (UVOT,XRT)},
\facility{CXO},
\facility{SOAR (Goodman)},
\facility{NTT (EFOSC2)},
\facility{CTIO: PROMPT}

\end{document}